\documentclass[prb, superscriptaddress]{revtex4}
\usepackage{epsfig}
\usepackage{color}
\usepackage{amsmath}
\usepackage{enumitem}
\usepackage{hyperref}
\hypersetup{
	colorlinks = true,
	linkcolor = blue,
	citecolor = blue
}

\usepackage{color}
\usepackage[normalem]{ulem}

\newcommand{\be}{\begin{equation}}
\newcommand{\ee}{\end{equation}}

\newcommand{\bea}{\begin{eqnarray}}
\newcommand{\eea}{\end{eqnarray}}

\newcommand{\p}{\partial}

\newcommand{\sgn}{{\rm sgn\,}}

\renewcommand{\vec}[1]{{\bf #1}}
\renewcommand{\phi}{\varphi}
\renewcommand{\epsilon}{\varepsilon}
\def\nn{\nonumber\\}

\begin{document}

\title{Electron states for gapped pseudospin-1 fermions in the field of charged impurity}
\date{\today}

\author{E. V. Gorbar}
\affiliation{Department of Physics, Taras Shevchenko National University of Kyiv, Kyiv, 03680, Ukraine}
\affiliation{Bogolyubov Institute for Theoretical Physics, Kyiv, 03680, Ukraine}

\author{V. P. Gusynin}
\affiliation{Bogolyubov Institute for Theoretical Physics, Kyiv, 03680, Ukraine}

\author{D. O. Oriekhov}
\affiliation{Department of Physics, Taras Shevchenko National University of Kyiv, Kyiv, 03680, Ukraine}

\begin{abstract}

The electron states of gapped pseudospin-1 fermions of the $\alpha-{\cal T}_3$ lattice in the Coulomb field of a charged impurity are studied.
The free $\alpha-{\cal T}_3$ model has three dispersive bands with two energy gaps between them depending on the parameter $\Theta$ which
controls the coupling of atoms of honeycomb lattice with atoms in the center of each hexagon, thus, interpolating between graphene $\Theta=0$
and the dice model $\Theta=\pi/4$. The middle band becomes flat one with zero energy in the dice model. The bound electron
states are found in the two cases: the centrally symmetric potential well and a regularized Coulomb potential of the
charged impurity. As the charge of impurity increases, bound state energy levels descend from the upper and central continua and dive
at certain critical charges into the central and lower continuum, respectively. In the dice model, it is found that the flat band
survives in the presence of a potential well, however, is absent in the case of the Coulomb potential. The analytical results are presented for
the energy levels near continuum boundaries in the
potential well. For the genuine Coulomb potential, we present the recursion relations that determine the coefficients of the series
expansion of wave functions of bound states. It is shown that the condition for the termination of the series expansion gives
two equations relating energy and charge values. Hence, analytical solutions can exist
for a countably infinite set of values of impurity charge at fixed $\Theta$.

\end{abstract}
\maketitle

\section{Introduction}

Like the electrons in graphene, low-energy gapless pseudospin-1 fermions have the energy spectrum linear
in momentum  except the additional flat band with zero energy [\onlinecite{Bercioux,Raoux}]. Completely flat bands
[\onlinecite{Heikkila}] can be realized in certain lattice models (for a recent review of artificial flat band
systems, see Ref.[\onlinecite{Flach}]). The $\alpha-{\cal T}_3$ lattice is one of the well-known realizations of pseudospin-1 fermions in two
dimensions (2D). It is a tight-binding model with atoms situated at both the vertices of a hexagonal lattice and the hexagons centers
[\onlinecite{Sutherland,Vidal}]. Since there are three sites per unit cell, the electron
states in the $\alpha-{\cal T}_3$ model are described by three-component fermions and the energy spectrum consists of three bands.
Experimentally, the $\alpha-{\cal T}_3$ lattice has been realized in Josephson arrays [\onlinecite{Serret}] and its optical realization by means
of laser beams was proposed in Ref.[\onlinecite{Rizzi}].

In the specific case when the two hopping amplitudes are equal, the $\alpha-{\cal T}_3$ lattice corresponds to the dice model. In linear order
to momentum deviations from the $K$ and $K^{\prime}$ points, the low-energy Hamiltonian of the dice model describes massless pseudospin-1
fermions and is given by the scalar product of momentum and the spin-1 matrices. The two energy bands form a Dirac cone and the third band is
completely flat and has zero energy [\onlinecite{Bercioux,Raoux}]. Several physical quantities have been studied in the
$\alpha-{\cal T}_3$ lattice such as orbital susceptibility [\onlinecite{Raoux}], optical conductivity [\onlinecite{Carbotte,Illes,Cserti}],
and magnetotransport [\onlinecite{Nicol,Biswas,Xu,Islam}].

Flat bands always attracted attention because quenching of the kinetic energy strongly enhances the role of electron-electron
interactions and may lead to a realization of many very interesting strongly correlated states like, for example, the fractional quantum Hall
states of 2D fermions. Sometime ago it was understood that flat bands could
be realized in twisted bilayer graphene [\onlinecite{Santos,Bistritzer}] and recently, indeed, the correlated insulator behavior at
half-filling [\onlinecite{Cao}] and unconventional superconductivity [\onlinecite{Fatemi}] was observed. The pairing problem in materials with
three bands crossing was studied in Ref.[\onlinecite{Lin}]. A pressure induced superconductivity was reported in MoP [\onlinecite{Chi}] which
hosts triply degenerate fermion states.

It is well known that the presence of boundaries and/or charged impurities removes the degeneracy of the electron states Landau levels. In a
recent paper [\onlinecite{dice-model-boundaries}], we showed that, remarkably, the energy dispersion of the completely flat
energy band of the dice model is not affected by the presence of boundaries except the trivial reduction of the degenerated electron states
due to the finite spatial size of the system. It was shown also that the flat band for dice lattice remains unaltered in the
presence of circularly polarized radiation [\onlinecite{Dey2018}]. The question whether the electron states of the flat band remain
degenerate in the presence of charged impurities provides a part of the motivation for the present study.

In order to gain an insight into the general response of three-component fermions to potential perturbations, we consider
the electron states of the $\alpha-{\cal T}_3$ lattice in two cases, namely, the radially symmetric potential well and a regularized
Coulomb potential, and determine how electron bound states evolve with the strength of potential splitting from and diving into the continuum
energy bands.

The paper is organized as follows. The $\alpha-{\cal T}_3$ model with a gap term is described in Sec.\ref{section:model}. The spectral equation
for pseudospin-1 fermions in a potential well is derived in Sec.\ref{sec:well}. The bound electron states on the $\alpha-{\cal T}_3$ lattice in
a potential well and the Coulomb field of a charged impurity are studied in Secs.\ref{sec:well-solutions} and \ref{section:center},
respectively. The obtained results are discussed and summarized in Sec.\ref{section:summary}. Some particular bound states
solutions in a potential well in the $\alpha-{\cal T}_3$ and dice models are given in Appendices \ref{sec:large-angle} and \ref{sec:dice}.

\section{Gapped $\alpha-{\cal T}_3$ model}
\label{section:model}

\begin{figure}[h!]
	\includegraphics[scale=0.4]{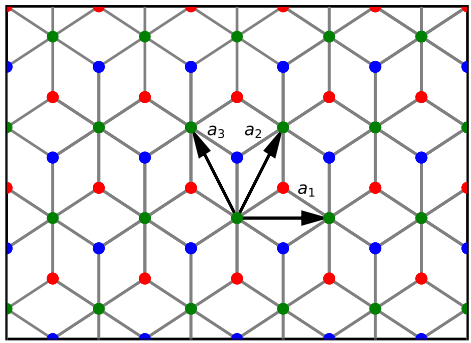}
	\caption{The $\alpha-{\cal T}_3$ lattice: red, blue and green points display atoms of the
		$A$, $B$ and $C$ sublattices, respectively. The vectors
		$\vec{a}_1=(\sqrt{3},\,0)a$ and $\vec{a}_2=(\sqrt{3}/2,\,3/2)a$ are the basis vectors of the $C$ sublattice and $a$ is
		the intersite distance.}
	\label{fig-lattice}
\end{figure}

The $\alpha-{\cal T}_3$ lattice has a unit elementary cell with three different lattice sites whose two sites (A,C) like in graphene
form a honeycomb lattice with hopping amplitude $t_{AC} = t_1$ and additional B sites at the center of each hexagon are
connected to the C sites with hopping amplitude $t_{BC} = t_2$ (see Fig.\ref{fig-lattice}). The two hopping parameters $t_1$ and $t_2$ are not equal in
general.
In this section, we will consider the low-energy free Hamiltonian of the gapped $\alpha-{\cal T}_3$ model, determine the Chern number of its
three bands, calculate the corresponding density of states, and write down the system of equations for eigenspinors of the model in a centrally
symmetric potential.

\subsection{Free Hamiltonian}

\begin{figure}[h!]
	\centering
	\includegraphics[scale=0.6]{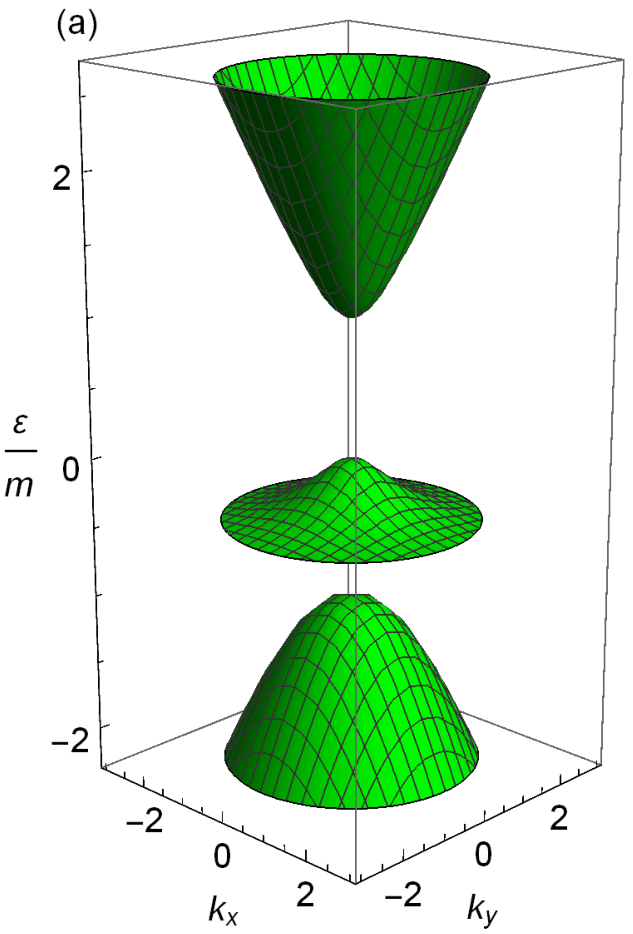}
	\includegraphics[scale=0.6]{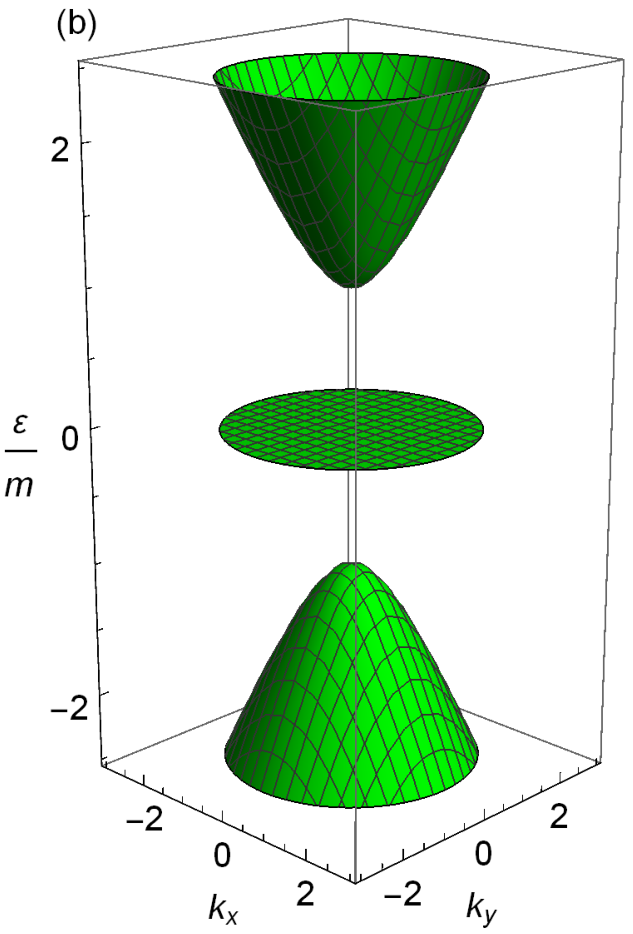}
	\includegraphics[scale=0.6]{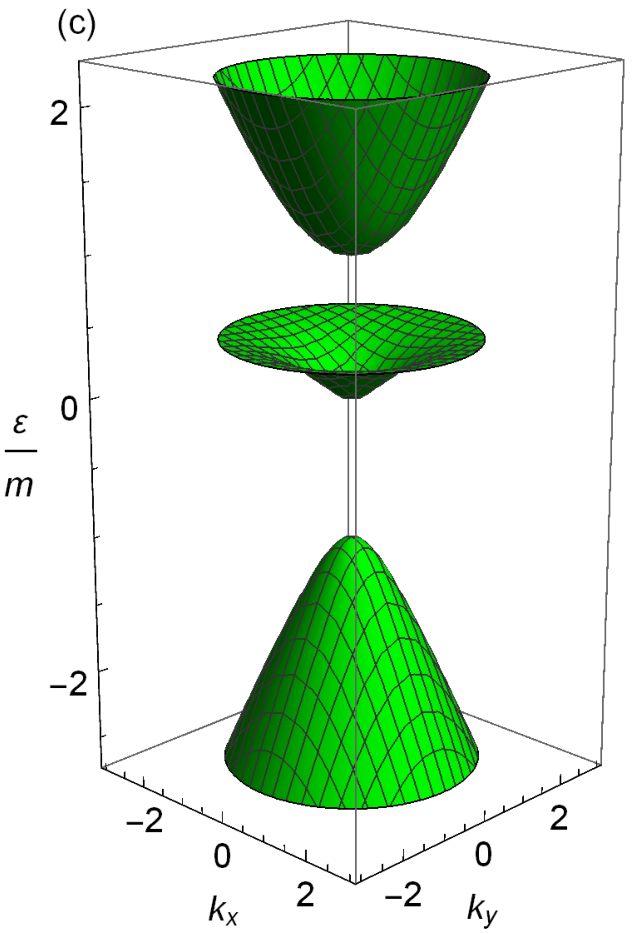}
	\includegraphics[scale=0.6]{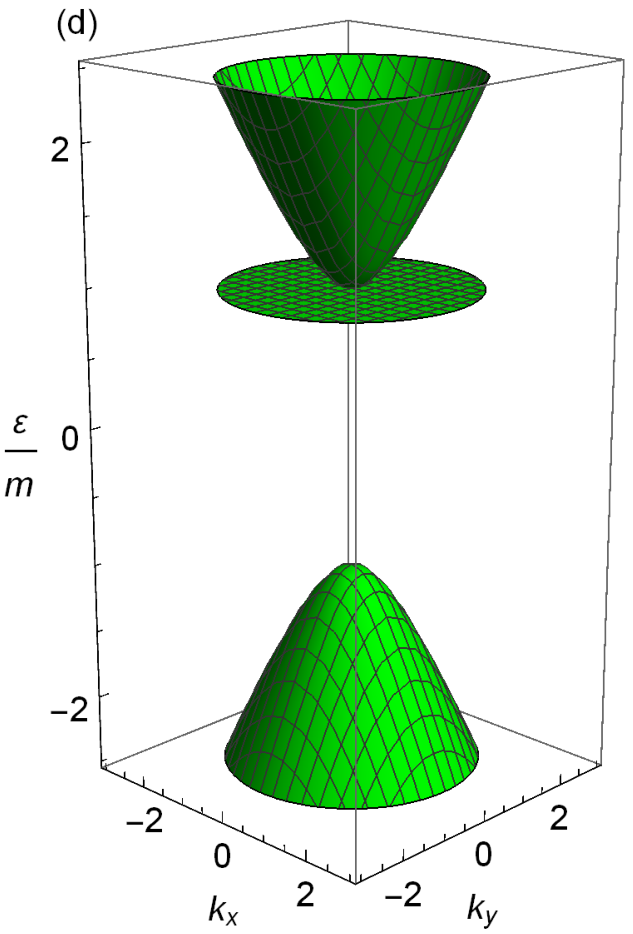}
	\caption{The band structure in momentum space of the $\alpha-\mathcal{T}_3$ model for $m=1$ and different values of angle $\Theta$:
	$\Theta=\frac{\pi}{6}$, $\Theta=\frac{\pi}{4}$ (dice model), and $\Theta=\frac{\pi}{3}$. In the panel (d): the band structure in case of the mass term $m \text{diag}(1,-1,1)$ with $m>0$.}
	\label{fig1}		
\end{figure}

We begin with the low-energy kinetic Hamiltonian of the $\alpha-{\cal T}_3$ model [\onlinecite{Raoux}] with an additional gap term
$m$ (we set $\hbar=v_F=1$)
\begin{align}
H_0(\vec{k},\xi)=\left(\begin{array}{ccc}
0 & \cos\Theta (\xi k_x-ik_y) & 0\\
\cos\Theta (\xi k_x+ik_y) & 0 & \sin\Theta(\xi k_x-ik_y)\\
0 & \sin\Theta(\xi k_x+ik_y) & 0
\end{array}\right) + m \left(\begin{array}{ccc}
1 & 0 & 0\\
0 & 0 & 0\\
0 & 0 & -1
\end{array}\right),
\label{Hamiltonian-free}
\end{align}
where $\tan\Theta=t_2/t_1$ and the valley index $\xi=\pm1$. The two valley Hamiltonian, $H_0(\vec{k},+1)\oplus H_0(\vec{k},-1)$,
is time-reversal invariant since under time-reversal  transformation $H^*_0(\vec{k},\xi)=H_0(-\vec{k},-\xi)$. In what follows we consider only
one valley Hamiltonian with $\xi=1$ for certainty, which breaks both the time reversal and particle-hole symmetries (the results for
the other valley can be readily obtained in a similar way). The one valley Hamiltonian can be realized, for  example, in the staggered-flux
kagome lattice [\onlinecite{Ohgushi,Green2010}].  The spectrum of the Hamiltonian can be found from the third order equation
\begin{align}
	\epsilon  (m^2 -\epsilon^2) +\left(k_x^2+k_y^2\right) (m  \cos (2 \Theta)+\epsilon )=0.
\end{align}
Since the coefficient near $\epsilon^2$ term is zero, three roots $\epsilon_{+},\epsilon_{-},\epsilon_{0}$ satisfy
	$\epsilon_{+}+\epsilon_{-}+\epsilon_{0}=0$.
In the particular case of the dice model $\Theta=\frac{\pi}{4}$, it is easy to determine the dispersion of energy bands
$\pm\sqrt{m^2+\mathbf{k}^2},\,0$ [\onlinecite{Green2010,Gao}]. The energy spectrum of the $\alpha-{\cal T}_3$ model is also easy to find in
the gapless case $m=0$ whose spectrum $\epsilon=\pm|\mathbf{k}|,0$ does not depend on $\Theta$
and the zero energy flat band touches two linearly dispersing bands at the point $\mathbf{k}=0$.
In the general case, the energy dispersion can be found by using the Cardano formulas
\be
\epsilon_\lambda(\mathbf{k})=2\sqrt{\frac{\mathbf{k}^2+m^2}{3}}\cos\left[\frac{1}{3}\arccos\left(\frac{3\sqrt{3}m\mathbf{k}^2
	\cos2\Theta}{2(\mathbf{k}^2+m^2)^{3/2}}\right)+\frac{2\pi(\lambda-1)}{3}\right],\quad \lambda=1,0,-1.
\label{dipersion}
\ee
At large values of $|\mathbf{k}|$ the middle band solution $\epsilon_{0}$ tends to $-m\cos(2\Theta)$. The other two bands are linear
at large momentum $\epsilon_{\pm}\simeq\pm|\mathbf{k}|$. For small momenta, $|\mathbf{k}|\ll m$, the bands behave as
\be
\epsilon_{\pm}(\mathbf{k})\simeq \pm m+\frac{\mathbf{k}^2(\pm1+\cos2\Theta)}{2m}, \quad \epsilon_{0}(\mathbf{k})
\simeq-\frac{\mathbf{k}^2\cos2\Theta}{m}.
\ee
Thus, for $\Theta\leq \frac{\pi}{4}$, there are two spectral gaps $\epsilon\in[0,m]$ and $\epsilon\in [-m,-m\cos(2\Theta)]$ and, for
$\Theta\geq \frac{\pi}{4}$, the spectral gaps are $\epsilon\in [-m,0]$ and $\epsilon\in[m|\cos(2\Theta)|, m]$.
For $\Theta\neq\frac{\pi}{4}$, the middle band $\epsilon_{0}(\mathbf{k})$ is also dispersive: it has a hole-like dispersion for
$0<\Theta<\frac{\pi}{4}$ and a electron-like for angles $\frac{\pi}{4}<\Theta<\frac{\pi}{2}$. One can switch between two possibilities
by changing hopping parameters in the Hamiltonian. The band structure for several values
of the angle $\Theta$ is shown in Fig.\ref{fig1}. For $\Theta=\pi/4$ (dice model), the middle band becomes dispersionless
and completely flat. The spectrum in this case is particle-hole symmetric with the isolated flat band separated from two dispersive bands
above and below by the gap $m$.
It is interesting that for another type of a gap term, $m \text{diag}(1,-1,1)$, the flat band with the energy $\epsilon=m$ exists for
arbitrary $\Theta$ and touches either the upper ($m>0$) or lower ($m<0$) dispersive energy band [\onlinecite{Raoux2015}]
(see panel (d) in Fig.\ref{fig1}).

The normalized spinor can be written in the form (up to overall phase)
 \bea
 \psi_\lambda(\mathbf{k})=N_\lambda(\mathbf{k})\hspace{-1mm}\left(\hspace{-1mm}\begin{array}{c}(m+\epsilon_\lambda)\cos\Theta (k_x-i k_y)\\
 \epsilon_\lambda^2-m^2\\ -(m-\epsilon_\lambda)\sin\Theta (k_x+i k_y)\end{array}\hspace{-1mm}\right)
 \hspace{-1mm},\quad N_\lambda(\mathbf{k})=\frac{1}{\sqrt{k^2 \left[m^2+\epsilon_\lambda^2 +2m\epsilon_\lambda \cos(2\Theta)\right]
+(m^2-\epsilon_\lambda^2)^2}}.
 \label{spinor-spin1}
 \eea
For the dice model ($\Theta=\pi/4$), it reduces to the expression in Appendix of Ref.[\onlinecite{Green2010}], while for $m=0$
and arbitrary $\Theta$ it reduces to the spinor in Ref.[\onlinecite{Raoux}] (see, also, Refs.[\onlinecite{Dey2018,Iurov}]).

\subsection{Berry phase and curvature}

The Berry phase is defined as a linear integral of the Berry connection
\be
\Phi_\lambda=\oint_\Gamma d\mathbf{k}{\cal A}^{\lambda}(\mathbf{k}),
\ee
where the Berry connection is given by the expression ${\cal A}^\lambda_i(\mathbf{k})=
-i\langle\psi_\lambda(\mathbf{k})|{\nabla}_{k_i}\psi_\lambda(\mathbf{k})\rangle$, and the integral goes over a close trajectory
around the K point with constant band energy $\epsilon_\lambda(\mathbf{k})$. The Berry phase is a gauge invariant quantity and its value
is unique up to $2\pi n$ with $n$ integer.
The Chern number of the $\lambda$ band is defined as an integral of the Berry curvature over the Brillouin zone
\be
C_\lambda=\frac{1}{2\pi}\int_{BZ} d^2k\,\Omega^{\lambda}(\mathbf{k}),
\label{Chern-number}
\ee
where $\Omega^\lambda(\mathbf{k})=(\pmb{\nabla}_k\times\mathbf{{\cal A}})_z$.  The Chern number is well defined when bands are
separated by gaps. Then the Chern number is topological and can only change when a band touching happens. In our low-energy theory
the integral in Eq.(\ref{Chern-number}) can be written as an integral at an infinitely distant circumference
\be
C_\lambda=\frac{1}{2\pi}\lim_{k\to\infty}\oint_\Gamma d\mathbf{k}{\cal A}^\lambda(\mathbf{k})=\frac{1}{2\pi}\lim_{k\to\infty}k\int\limits_0^{2\pi}
d\varphi {\cal A}^\lambda_\varphi(\mathbf{k}),\quad k=|\mathbf{k}|,
\ee
where the angular component of the connection is calculated as
\be
{\cal A}^\lambda_\varphi(\mathbf{k})=-\frac{i}{k}\langle\psi_\lambda(\mathbf{k})|\frac{\partial}{\partial\varphi}\psi_\lambda(\mathbf{k})\rangle.
\ee

The middle component of spinor (\ref{spinor-spin1}) is real, however, it vanishes for $\lambda=\pm$ at $\mathbf{k}=0$ where
$\epsilon_{\pm}(\mathbf{k}=0)=\pm m$. Therefore, its phase cannot be fixed at this point. For definiteness, let us consider the upper band
$\lambda=+$ (the analysis in the case of the lower band $\lambda=-$ is quite similar). Since $N_\lambda(\mathbf{k})\sim 1/k$ as $k \to 0$,
the upper component of the spinor (\ref{spinor-spin1}) for $m>0$ does not vanish at $\mathbf{k}=0$. Obviously, we can make it real by
multiplying the spinor (\ref{spinor-spin1}) by $\mbox{exp}[i\phi]$ where $\varphi=\arctan(k_y/k_x)$. [If $m<0$, then the third component
of the spinor does not vanish at $\mathbf{k}=0$ and can be made real multiplying the spinor by $\mbox{exp}[-i\phi]$. The subsequent analysis is
similar to the case $m>0$.] We obtain the following spinor:
\bea
 \psi_+ (\mathbf{k})= N_+(\mathbf{k})\left(\begin{array}{c}(m+\epsilon_+)k\cos\Theta\\(\epsilon_+^2-m^2)e^{i\phi}\\
 -(m-\epsilon_+)k\sin\Theta\,e^{2i\phi}
 \end{array}\right).
 \label{spinor-spin1-real}
 \eea
Then we find for arbitrary $m$,
\be
{\cal A}^+_{\phi}(\mathbf{k})=\frac{\mbox{sgn}(m)}{k}\frac{(m^2-\epsilon_+^2)^2+2k^2(|m|-\epsilon_+)^2\sin^2\Theta}{k^2 \left[m^2+\epsilon_+^2
+2m\epsilon_+ \cos(2\Theta)\right]+(m^2-\epsilon_+^2)^2}.
\ee
Combining the results for all three bands and calculating the limit $k\to\infty$, we find the Berry phases for the $K$ point ($\xi=1$)
\be
\Phi_{\lambda=\pm1}=\lim_{k\to\infty}\oint_\Gamma d\mathbf{k}{\cal A}^\pm(\mathbf{k})=2\pi(\lambda\mbox{sgn}(m)-\frac{\cos2\Theta}{2}),\quad \Phi_{\lambda=0}=2\pi\cos2\Theta,
\ee
which are in agreement with Ref.[\onlinecite{Raoux}] for $\Theta=\pi/4$ in the gapless dice model (we assume $\mbox{sgn}(0)=0$) except
the overall sign minus (due to the definition of the connection with opposite sign). The Chern numbers $C_\lambda=\Phi_\lambda/2\pi$ are
\be
 C_{\lambda=\pm1}=\lambda\,\mbox{sgn}(m)-\frac{1}{2}\cos2\Theta,\quad C_{\lambda=0}=\cos2\Theta.
\label{Chern-number-final}
\ee
For the gapped dice model ($\Theta=\pi/4$), the Chern numbers coincide with those in Ref.~[\onlinecite{Green2010}].
Note that $\sum_{\lambda}C_\lambda=0$. The Berry phases for the $K'$ point ($\xi=-1$) have opposite signs.
As it was noticed in Ref.[\onlinecite{Raoux}] the Berry phases are topological but not $\pi$ quantized. Therefore, the corresponding
Chern numbers calculated in an effective low energy theory are non-integer in general. The situation is similar to the
calculation of the Chern number at the Dirac point in 2D, whose Chern number is half-integer. As discussed in Ref.[\onlinecite{Volovik}], this
happens because the momentum space is not compact and the unit vector which characterizes the mapping from the momentum space to the Hamiltonian
does not tend to the same value at infinity, but instead forms the 2D hedgehog. Therefore, it is better and more appropriate to
characterize the considered low-energy theory through fluxes of the Berry curvature rather than the Chern number.

\begin{figure}[h]
	\centering
	\includegraphics[scale=0.6]{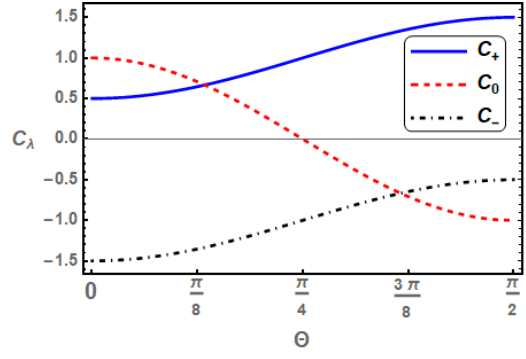}
	\caption{The Chern numbers as a function of the parameter $\Theta$ for $m>0$.}
	\label{fig-Chern}		
\end{figure}

The Chern numbers can be calculated also by using the formula for the Berry curvature written as a sum
over the eigenstates [\onlinecite{XiaoRMP2010}],
\be
\Omega^\lambda(\mathbf{k})=-i\sum_{\lambda'\neq\lambda}\frac{\langle\psi_\lambda|\partial H(\mathbf{k})/
\partial k_x|\psi_{\lambda'}\rangle \langle\psi_{\lambda'}|\partial H(\mathbf{k})/
\partial k_y|\psi_{\lambda}\rangle-(k_x\leftrightarrow k_y)}{(\epsilon_\lambda-\epsilon_{\lambda'})^2}.
\label{Bcurvature-over-bandsspinors}
\ee
This formula is manifestly gauge independent and has the advantage that no differentiation of the spinor functions is involved, therefore it can be
evaluated for any fixed gauge. Hence it is no longer necessary to pick spinors to be smooth and single valued.  Also this formula suggests that
the Berry curvature could be divergent when there is a band touching as it contains $(\epsilon_\lambda-\epsilon_{\lambda'})^2$ in the
denominator. The results of numerical calculations of the Chern numbers are shown in Fig.\ref{fig-Chern} and are in agreement with the analytical
formulas (\ref{Chern-number-final}) for $\Theta\neq0,\pi/2$. The angles $\Theta=0,\pi/2$ should be considered separately since in this case hoppings
either from atoms A or atoms B vanish and we deal with graphene-like model for which we find $C=\pm 1/2$.
We note that the Chern numbers are, of course, integer in a microscopic theory where the Brillouin zone has the form of a torus (see, for
example, Ref.[\onlinecite{Ohgushi}] for a similar model.)

\subsection{Density of states}

\begin{figure}
	\centering
	\includegraphics[scale=0.33]{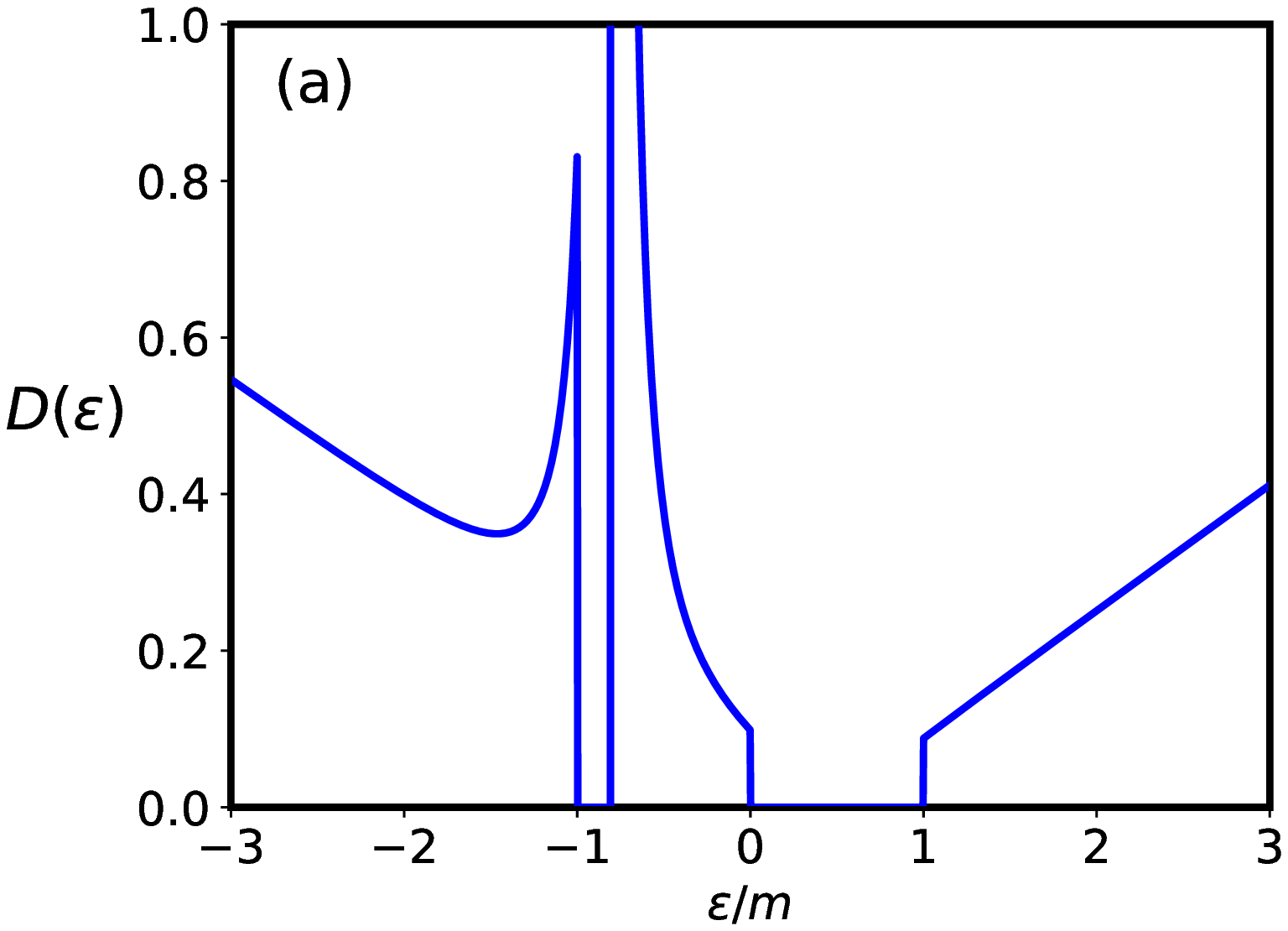}
	\includegraphics[scale=0.33]{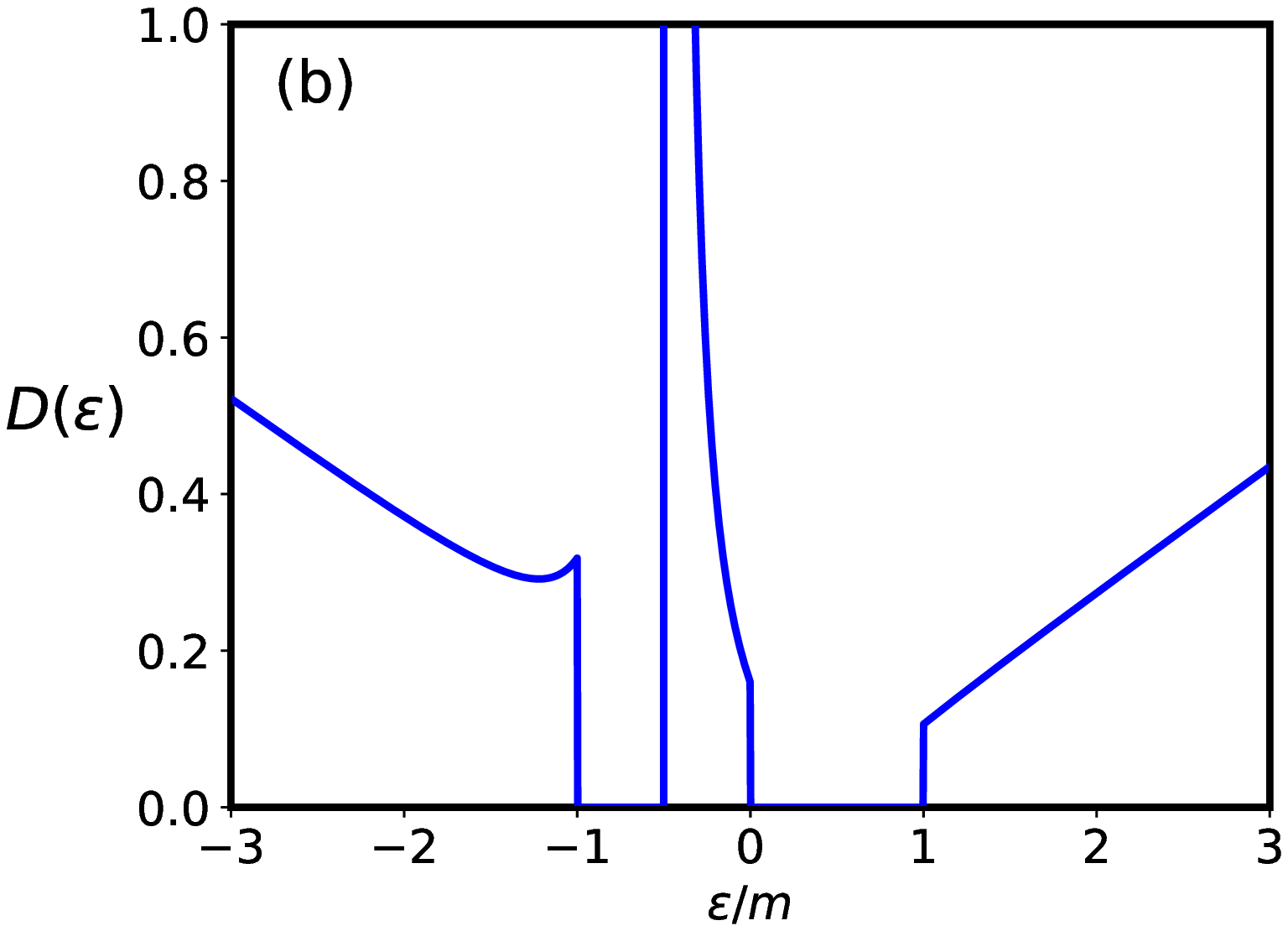}
	\includegraphics[scale=0.33]{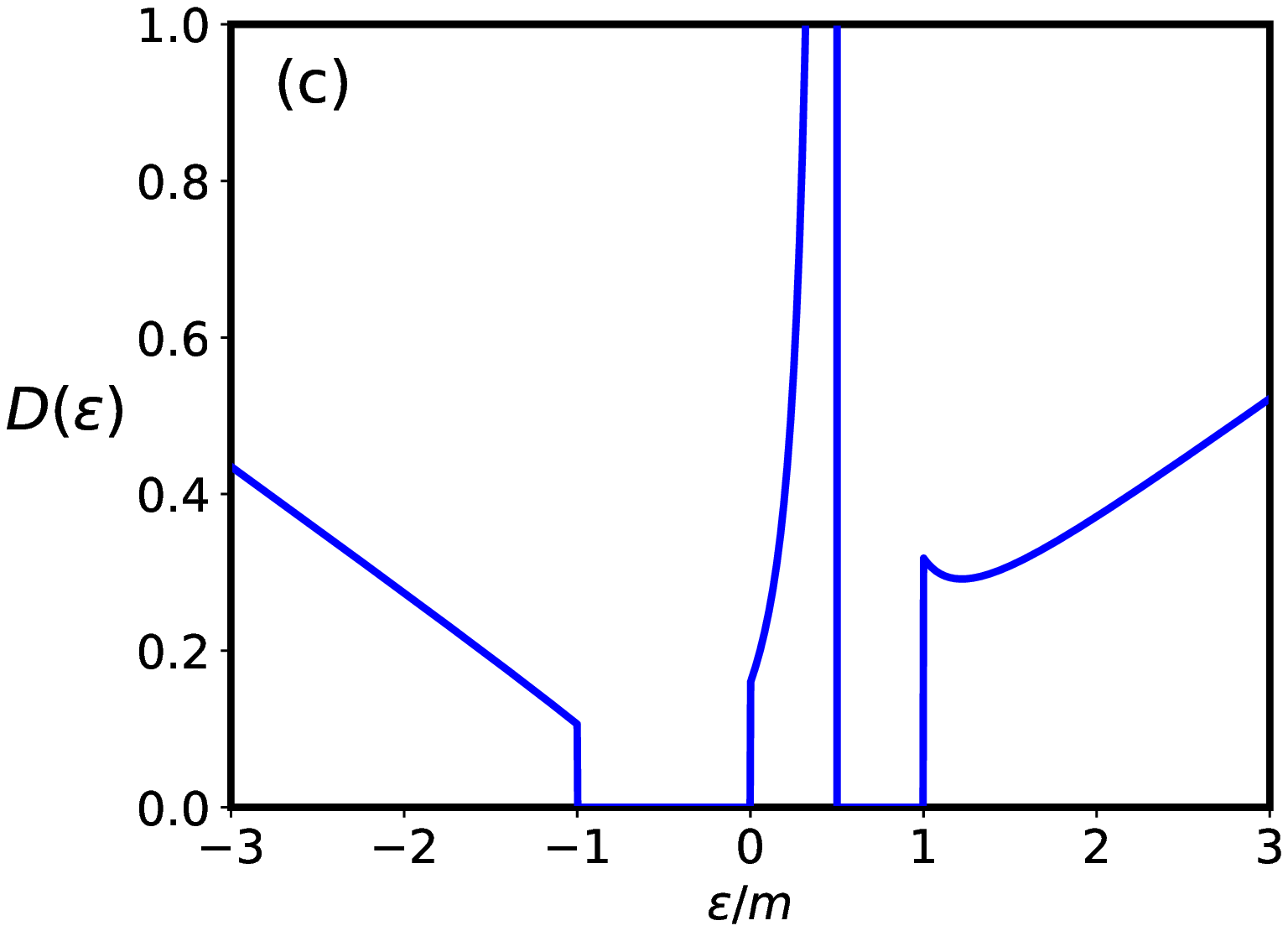}
	\caption{The density of states \eqref{eq:dos} for $m=1$ and for angles $\Theta=\frac{\pi}{10}$, $\Theta=\frac{\pi}{6}$, $\Theta=\frac{\pi}{3}$.}
	\label{fig:dos}		
\end{figure}

By definition, the two-dimensional density of states (DOS) is given by
\begin{align}
D(\epsilon)=\sum\limits_{\lambda=0,\pm1}\int \frac{d^2 k}{(2\pi)^2}\delta(\epsilon-\epsilon_{\lambda}(\vec{k}))
=\sum\limits_{\lambda=0,\pm1}\int_{0}^{\infty}\frac{dk^2}{4\pi}\delta(\epsilon-\epsilon_{\lambda}(\mathbf{k}^2)).
\label{eq:dos}
\end{align}
In the particular case of the dice model, we can calculate the DOS analytically. The energy dispersion is given by
$\epsilon_{\lambda}(k)=\lambda\sqrt{m^2+k^2}, 0$ and the density of states equals
\begin{align}
D(\epsilon)=\begin{cases}
\frac{|\epsilon|}{2\pi}, &|\epsilon|\ge m,\\
\delta(\epsilon) \Omega_{k}, & |\epsilon|< m,
\end{cases}
\end{align}
$\Omega_{k}=\int_{0}^{k_{max}}\frac{dk^2}{4\pi}$ is the volume in momentum space.
For other values of $\Theta$, the DOS cannot be found analytically.  The
results of numerical computations are shown in Fig.\ref{fig:dos} for the three values of $\Theta=\pi/10,\pi/6, \pi/3$.

\subsection{Hamiltonian with centrally symmetric potential}

The low-energy Hamiltonian of the gapped $\alpha-\mathcal{T}_{3}$ lattice model [\onlinecite{Raoux}] for pseudospin-1
fermions in a centrally symmetric potential $V(r)$ reads in the polar coordinate system
\begin{align}
H=\left(\begin{array}{ccc}
m & -i\cos\Theta e^{-i\phi}(\p_r-\frac{i}{r}\p_{\phi}) & 0\\
-i\cos\Theta e^{i\phi}(\p_r+\frac{i}{r}\p_{\phi}) & 0 & -i\sin\Theta e^{-i\phi}(\p_r-\frac{i}{r}\p_{\phi})\\
0 & -i\sin\Theta e^{i\phi}(\p_r+\frac{i}{r}\p_{\phi}) & -m
\end{array}\right) + V(r).
\end{align}
The total angular momentum operator is $J=L_z+\hbar S_z=-i\p_\phi+ S_z$ (the matrix $S_z$ is the matrix near $m$ in
Eq.(\ref{Hamiltonian-free})). We seek the spinor as an eigenvector of $J$ with quantum number $j$ in the form
\begin{equation}
	\Psi=\frac{1}{r}\left(\begin{array}{c}
	a(r) e^{i(j-1)\phi}\\
	ic(r) e^{ij\phi}\\
	b(r) e^{i(j+1)\phi}
	\end{array}\right)
\label{spinor-general-form-polar-coordinates}
\end{equation}
and obtain the system of equations for components
\begin{align}
\label{eq:main-system-1}
	&\cos\Theta\left(c'+\frac{j-1}{r}c\right)+\left(m+V(r)-\epsilon\right)a=0,\\
\label{eq:main-system-2}	
	&\cos\Theta\left(a'-\frac{j}{r}a\right)+\sin\Theta\left(b'+\frac{j}{r}b\right)+\left(-V(r)+\epsilon\right)c=0,\\
\label{eq:main-system-3}
	&\sin\Theta\left(c'-\frac{j+1}{r}c\right)+\left(-m+V(r)-\epsilon\right)b=0.
\end{align}
Combining Eqs.(\ref{eq:main-system-1}) and (\ref{eq:main-system-3}), it is easy to find that
the system above gives the linear relation between three components ($\Theta\neq0,\pi/2$)
\begin{align}
	\frac{2j}{r}c+\frac{m-\epsilon+V(r)}{\cos\Theta}a+\frac{m+\epsilon-V(r)}{\sin\Theta}b=0,
\label{constraint-equation}
\end{align}
which allows one to reduce system (\ref{eq:main-system-1})-(\ref{eq:main-system-3}) to a second order differential equation for one
component (see, e.g., Eq.(\ref{dice-model-equation-c}) below found in the case of a potential well).
We will analyze the solutions of the above system of equations in the simplest setting of a potential well, as well as
the Coulomb centre.

\section{Potential well: spectral equation}
\label{sec:well}

In this section we consider the case of the potential well with the potential $-V_0\Theta(r_0-r)$, $V_0>0$.
%
Since the potential does not depend on $r$ for $r<r_0$ and $r>r_0$, we can obtain the following second-order equation for $c(r)$:
\be
\frac{m\cos(2\Theta)+\epsilon+V_0\Theta(r_0-r)}{m^2-(\epsilon+V_0\Theta(r_0-r))^2}\left(c''-\frac{1}{r}c'-\frac{j^2-1}{r^2}c\right)
-(\epsilon+V_0\Theta(r_0-r)) c=0.
\label{dice-model-equation-c}
\ee
We are interested in bound states solutions. We search them in the form
\begin{align}
	c(r)=\begin{cases}
	c_1 r J_{|j|}(v_1 (\epsilon+V_0)r)+\delta_{j,0}\tilde{c}_1 r Y_0(v_1(\epsilon+V_0) r) ,& r<r_0,\\
	c_2 r K_{|j|}(v_2 r),& r>r_0,
	\end{cases}
\label{c-component-well}
\end{align}
where we use the notation
\begin{align}
	v_1(\epsilon)=\sqrt{\frac{\epsilon (\epsilon^2-m^2)}{\epsilon+m \cos (2 \Theta )}},\quad v_2(\epsilon)
	=\sqrt{\frac{\epsilon (m^2-\epsilon^2)}{\epsilon+m \cos (2 \Theta )}}.
\end{align}

By using Eqs.(\ref{c-component-well}) and (\ref{eq:main-system-1}), we find the $a$ component
\begin{align}
   a=\begin{cases}
   -\frac{\cos\Theta}{m-\epsilon-V_0}\sgn(j)v_1 (\epsilon+V_0)r\bigg(c_1  J_{|j|-\sgn(j)}(v_1(\epsilon+V_0)r)+\tilde{c}_1 \delta_{j,0}
   Y_{-1}(v_1(\epsilon+V_0)r) \bigg),
   & r<r_0,\\
   \frac{\cos\Theta}{m-\epsilon} c_2 v_2 r K_{|j|-\sgn(j)}(v_2r), & r>r_0.
   \end{cases}
\label{a-component-well}
\end{align}
In this function we have $Y_{-1}(r)$, which is singular at origin and can not meet normalization condition. Therefore, we set $\tilde{c}_1=0$.
Similarly, for the $b$ component, we get
\begin{align}
   b=\begin{cases}
   -\frac{\sin\Theta}{m+\epsilon+V_0}{\rm sgn}(j)v_1 (\epsilon+V_0)r\bigg(c_1  J_{|j|+{\rm sgn}(j)}(v_1(\epsilon+V_0)r)  \bigg),& r<r_0,\\
   -\frac{\sin\Theta}{m+\epsilon} c_2 v_2 r K_{|j|+{\rm sgn}(j)}(v_2r), & r>r_0,
   \end{cases}
\label{b-component-well}
\end{align}
where we used the formulas
\begin{align}
&z J'_{|j|}(z)\pm j J_{|j|}(z)=\pm{\rm sgn}(j)z J_{|j|\mp{\rm sgn}(j)}(z)=\left\{\begin{array}{c}
\pm z J_{|j|\mp1}(z), \quad j\ge0,\\
\mp z J_{|j|\pm 1}(z), \quad j \,\le0,\end{array}\right.
\label{relations-BesselJ}
\end{align}
\begin{align}
z K'_{|j|}(z)\pm j K_{|j|}(z)=-z K_{|j|\mp{\rm sgn}(j)}(z)=\left\{\begin{array}{c}
-z K_{|j|\mp1}(z), \quad j\ge0,\\
-z K_{|j|\pm1}(z), \quad j \,\le0.\end{array}\right.
\label{relations-BesselK}
\end{align}

\subsection{Matching conditions and spectral equation}

Integrating the system of equations \eqref{eq:main-system-1}-\eqref{eq:main-system-3} for $V(r)=-V_0\theta(r_0-r)$ from $r_0-\delta/2$ up to
$r_0+\delta/2$ and then taking the limit $\delta \to 0$, we find that the 1st and 3rd equations imply that the $c$-component is continuous at $r=r_0$, i.e.,
\begin{equation}
c_>=c_<\,.
\label{EOM-1-condition}
\end{equation}
The 2nd equation of the system gives, obviously, the second matching condition
\begin{equation}
\cos\Theta\,a_> + \sin\Theta\,b_> = \cos\Theta\,a_< + \sin\Theta\,b_<.
\label{EOM-2-condition}
\end{equation}
Notice that there are only two matching conditions that is consistent with the fact that the system under consideration
of three first-order differential equations for the three components reduces to the second-order differential equation for one component
(\ref{dice-model-equation-c}).

By using Eqs.(\ref{c-component-well}), (\ref{a-component-well}), and (\ref{b-component-well}), we obtain that the matching conditions
(\ref{EOM-1-condition}) and (\ref{EOM-2-condition}) lead to the following spectral equation for bound states:
\begin{eqnarray}
&&\frac{\cos^2\Theta v_1(\epsilon+V_0)}{m-\epsilon-V_0}\frac{\sgn(j)J_{|j|-\sgn(j)}(v_1(\epsilon+V_0)r_0)}{J_{|j|}(v_1 (\epsilon+V_0)r_0)}
+\frac{\sin^2\Theta v_1(\epsilon+V_0)}{m+\epsilon+V_0}\frac{\rm{sgn}(j)J_{|j|+\rm{sgn}(j)}(v_1(\epsilon+V_0)r_0)}
{J_{|j|}(v_1 (\epsilon+V_0)r_0)}=\nonumber\\
&&-\frac{\cos^2\Theta v_2(\epsilon)}{m-\epsilon}\frac{K_{|j|-\sgn(j)}(v_2(\epsilon)r_0)}{K_{|j|}(v_2(\epsilon)r_0)}
+\frac{\sin^2\Theta v_2(\epsilon)}{m+\epsilon}\frac{K_{|j|+\rm{sgn}(j)}(v_2(\epsilon)r_0)}{K_{|j|}(v_2(\epsilon)r_0)}.
\label{spectrum-1}
\end{eqnarray}

The centrifugal barrier grows with $j$, therefore, we will consider in our analytic results only the electron bound states with $j=0$ and $j=1$
angular momenta. For $j=0$, the above spectral equation takes the form
\be
\frac{\epsilon+V_0+m\cos2\Theta}{(\epsilon+V_0)^2-m^2}\frac{v_1(\epsilon+V_0)J_{1}(v_1(\epsilon+V_0)r_0)}{J_{0}(v_1 (\epsilon+V_0)r_0)}
=\frac{\epsilon+m\cos2\Theta}{\epsilon^2-m^2}\frac{v_2(\epsilon)K_{1}(v_2(\epsilon)r_0)}{K_{0}(v_2(\epsilon)r_0)}.
\label{spectrum-2a}
\ee
In the case $j=1$, Eq.\eqref{spectrum-1} becomes
\begin{eqnarray}
&&\left(\frac{\cos^2\Theta }{m-\epsilon-V_0}J_{0}(v_1(\epsilon+V_0)r_0)
+\frac{\sin^2\Theta }{m+\epsilon+V_0}J_{2}(v_1(\epsilon+V_0)r_0)
\right)\frac{v_1(\epsilon+V_0)}{J_{1}(v_1 (\epsilon+V_0)r_0)}=\nonumber\\
&&\left(-\frac{\cos^2\Theta }{m-\epsilon}K_{0}(v_2(\epsilon)r_0)
+\frac{\sin^2\Theta v_2(\epsilon)}{m+\epsilon}K_{2}(v_2(\epsilon)r_0)\right)\frac{v_2(\epsilon)}{K_{1}(v_2(\epsilon)r_0)}.
\label{spectral-equation-j1}
\end{eqnarray}

\section{Potential well: solutions for $\Theta < \pi/4$}
\label{sec:well-solutions}

In this section, we will determine bound state solutions for pseudospin-1 fermions in the $\alpha-{\cal T}_3$ model in a potential
well for $0<\Theta < \pi/4$. Solutions for $\Theta>\frac{\pi}{4}$ and for $\Theta=\pi/4$ in the dice model can be found in a similar way and
the details of the corresponding analysis are given in Appendices \ref{sec:large-angle} and
\ref{sec:dice}. Since the states with total angular momenta $j=0$ and $j=1$ are the lowest energy states, only these states will be
analyzed explicitly.

\subsection{Solutions descending from the upper continuum}
\vspace{3mm}

Since $\epsilon \to m$ near the upper continuum, we have
\be
v_2(\epsilon)\simeq\sqrt{\frac{2m(m-\epsilon)}{1+\cos2\Theta}}\rightarrow0,
\ee
and we can use the asymptotes of the MacDonald functions
\begin{equation}
K_0(z) \approx - \ln\frac{z e^\gamma}{2}+O\left(z^2\ln z\right),\quad K_1(z) \approx \frac{1}{z}+O\left(z\ln z\right),
\quad K_2(z) \approx \frac{2}{z^2}-\frac{1}{2}+O\left(z^2\ln z\right),
\end{equation}
on the right-hand side of Eq.(\ref{spectrum-1}). For the state $j=0$, Eq.(\ref{spectrum-2a}) gives
\be
\frac{V_0+m(1+\cos2\Theta)}{V_0(2m+V_0)}\frac{v_1(m+V_0)J_1(v_1(m+V_0)r_0)}{J_0(v_1(m+V_0)r_0)}=-\frac{1+\cos2\Theta}{2(m-\epsilon)}
\ln^{-1}\left(\frac{2e^{-\gamma}}{r_0}\sqrt{\frac{1+\cos2\Theta}{2m(m-\epsilon)}}\right).
\label{eq:V0crit}
\ee
Clearly, this equation does not have solution at small potential strength, $V_0 \to 0$, since the left hand side of the equation is positive
while the right hand side is negative. A nontrivial solution appears when $V_0$ exceeds some critical value $V_{0,cr}$ which is determined
by the first zero $z=j_{0,1}\approx2.4$ of the Bessel function $J_0(z)$. This gives the equation for the critical potential
$v_1(m+V_{0,cr})r_0=j_{0,1}$. For $V_0-V_{0,cr}\ll V_{0,cr}$ taking into account that $J_0(v_1(m+V_0)r_0)\simeq
-J_1(j_{0,1})v'_1(m+V_{0,cr})r_0(V_0-V_{0,cr})$, Eq.(\ref{eq:V0crit}) simplifies to
\be
(m-\epsilon)\ln\frac{a}{m-\epsilon}=b(V_{0,cr})(V_0-V_{0,cr}),
\ee
where
\be
 a=\frac{2e^{-2\gamma}(1+\cos2\Theta)}{mr_0^2},\quad b(V_{0,cr})=
\frac{(1+\cos2\Theta)V_{0,cr}(2m+V_{0,cr})}{V_{0,cr}+m(1+\cos2\Theta)}\frac{v'_1(m+V_{0,cr})}{v_1(m+V_{0,cr})}.
\ee
The solution of this equation is given
\be
\epsilon=m- a\exp W_{-1}\left(-\frac{b(V_{0,cr})(V_0-V_{0,cr})}{a}\right),
\ee
where $W_{-1}(x)$ is a non principal branch of the Lambert function which is real for $- 1/e <x<0$ [\onlinecite{Corless}]. Since
\be
W_{-1}(-x)\simeq \ln x -\ln(-\ln x)\quad\quad \mbox{for} \quad x\to0,
\ee
we find
\be
\epsilon\simeq m+\frac{b(V_{0,cr})(V_0-V_{0,cr})}{\ln\frac{b(V_{0,cr})(V_0-V_{0,cr})}{a}}, \quad\quad
b(V_{0,cr})\simeq\left\{\begin{array}{c}1,\quad V_{0,cr}\ll m,\\ 1+\cos2\Theta,\quad V_{0,cr}\gg m.\end{array}\right.
\ee

For {\bf the state $j=1$}, Eq.(\ref{spectrum-1}) takes the following form near the upper continuum:
\be
-\frac{\cos^2\Theta v_1(m+V_0)}{V_0}\frac{J_{0}(v_1(m+V_0)r_0)}{J_{1}(v_1 (m+V_0)r_0)}
+\frac{\sin^2\Theta v_1(m+V_0)}{2m+V_0}\frac{J_{2}(v_1(m+V_0)r_0)}{J_{1}(v_1 (m+V_0)r_0)}=-m r_0\ln\frac{2e^{-\gamma}}{v_2(\epsilon\to m)r_0}
+\frac{\sin^2\Theta}{m r_0}.
\ee
This equation has the solution at any small $V_0$
\be
\epsilon=m\left\{1-\frac{2(1+\cos\,2\Theta)}{(mr_0)^2}\,\exp\left[-(1+\cos2\Theta)\left(\frac{2}{m V_0r^2_0}
+\frac{\tan^2\Theta}{m^2r^2_0}\right)-2\gamma\right]\right\}.
\ee
For $\Theta=0$, the above expression is similar to that in graphene (see Eq.(2.19) in Ref.[\onlinecite{Gamayun2011}]). While the
spectral gap in graphene is between $-\Delta/\hbar v_F$ and $\Delta/\hbar v_F$, the spectral gap here is between $0$ and $m$.
Therefore, in order obtain the same formula as in graphene one should subtract $m/2$ from $\epsilon$ and then replace $m/2\to\Delta/\hbar v_F$.

\subsection{Other analytical solutions}

Let us consider solutions diving into the central continuum. For the state $j=0$ and $\epsilon \to 0$,
the critical potential $V_{0,cr}$ is defined by the equation
\be
v_1(V_{0,cr})r_0=\sqrt{\frac{V_{0,cr}(V_{0,cr}^2-m^2)}{V_{0,cr}+m\cos2\Theta}}r_0= j_{1,1}\approx3.832
\label{V0crit-j0-nearepsilon0}
\ee
with the energy dispersion for $V_0\lesssim V_{0,cr}$ given by
\be
\epsilon=\frac{\cos2\Theta}{m r_0^2}\exp\left(-2\frac{\cos2\Theta}{am r_0^2(V_{0,cr}-V_0)}-2\gamma\right).
\ee
For the $j=1$ state, the critical potential is determined by the transcendental equation
	\begin{eqnarray}
	\left(\frac{\cos^2\Theta }{m-V_0}J_{0}(v_1(V_0)r_0)+\frac{\sin^2\Theta }{m+V_0}J_{2}(v_1(V_0)r_0)\right)
	\frac{v_1(V_0)}{J_{1}(v_1 (V_0)r_0)}=\frac{2\sin^2\Theta}{m r_0}.
	\end{eqnarray}
We can find the energy of bound states for near critical potentials expanding Eq.\eqref{spectral-equation-j1} to the linear order in $\epsilon$
and $V_{0}-V_{0,cr}$. As a result, we obtain the Lambert-type equation, which has the solution
 	\begin{align}
 	\epsilon = -\frac{2D(V_{0,cr})(V_{0,cr}-V_0)}{r \ln\left(\frac{2D(V_{0,cr})(V_{0,cr}-V_0)}{r a}\right)},\quad
 	a=\exp\left\{\frac{2D(V_{0,cr})}{r_0}\right\} \left[e^{2\gamma} \frac{mr_0^2}{4\cos(2\Theta)}\right]^{-1}.
 	\end{align}
Here $D(V_0)$ is the short-hand notation for the derivative of the left-hand side of Eq.\eqref{spectral-equation-j1} with respect to $V_0$ at
$\epsilon=0$.

The spectral equation (\ref{spectrum-2a}) for the state $j=0$ splitting from the middle band takes the form
\begin{align}
\frac{V_0-x}{(V_0-m\cos 2\Theta-x)^2-m^2}\frac{v_1(V_0-m\cos 2\Theta-x)J_{1}(v_1(V_0-m\cos 2\Theta -x)r_0)}{J_{0}(v_1 (V_0-m\cos 2\Theta-x)r_0)}
=\frac{\epsilon+m\cos2\Theta}{m^2-\epsilon^2}\frac{v_2(\epsilon)K_{1}(v_2(\epsilon)r_0)}{K_{0}(v_2(\epsilon)r_0)}.
\end{align}
For $V_0<x$, while the left-hand side is always positive, the right-hand side is negative, therefore, there are no solutions in this case.
When $V_0\to x$ from the above, the left-hand side oscillates in the range $(-\infty,\infty)$, while the right-hand side remains constant at
a given $x$. Therefore, there are many solutions when $V_0-x\to +0$ reflecting an essential singularity in the Bessel functions at
large value of their argument. The same situation takes place for each $j$, therefore, there are always the corresponding solutions for
$V_0\to 0$.

Finally, let us consider states diving into the lower continuum. They exist for $V_0$ in the interval
$m(1-\cos\,2\Theta) < V_0 < m$. The critical value $V_{0,cr}$ for the state $j=0$ is defined by the equation
\be
\frac{V_0-m(1-\cos2\Theta)}{V_0(2m-V_0)}\frac{v_1(-m+V_{0,cr})}{r_0|v^{\prime}_1(-m+V_{0,cr})|(V_{0,cr}-V_0)}
=\frac{1-\cos2\Theta}{2x r_0}\ln^{-1}\frac{2e^{-\gamma}}{v_2(-m+x)r_0}.
\label{diving-j0-2}
\ee
The critical value $V_{0,cr}$ when the $j=1$ states dives into the lower continuum is defined by the equation $J_1(v_1(-m+V_{0,cr})r_0)=0$,
i.e., the first zero of the Bessel function $J_1(z)$: $v_1(-m+V_{0,cr})r_0=j_{1,1}\approx3.832$.
It is not difficult to check that $V_{0,cr}$ for
diving into the lower continuum $\epsilon=-m$ of the state $j=1$ is less than $V_{0,cr}$ for the
state $j=0$. According to Eq.(\ref{relations-BesselJ}), $J_2(j_{1,1})=-J_0(j_{1,1})$, and $J'_1(j_{1,1})=J_0(j_{1,1})$. The corresponding
solution near $V_{0,cr}$ behaves as $\epsilon\sim-m +a(V_{0,cr}-V_0)$ with $a>0$.

\subsection{Numerical solutions: general picture}

\begin{figure}[h!]
	\centering
	\includegraphics[scale=0.45]{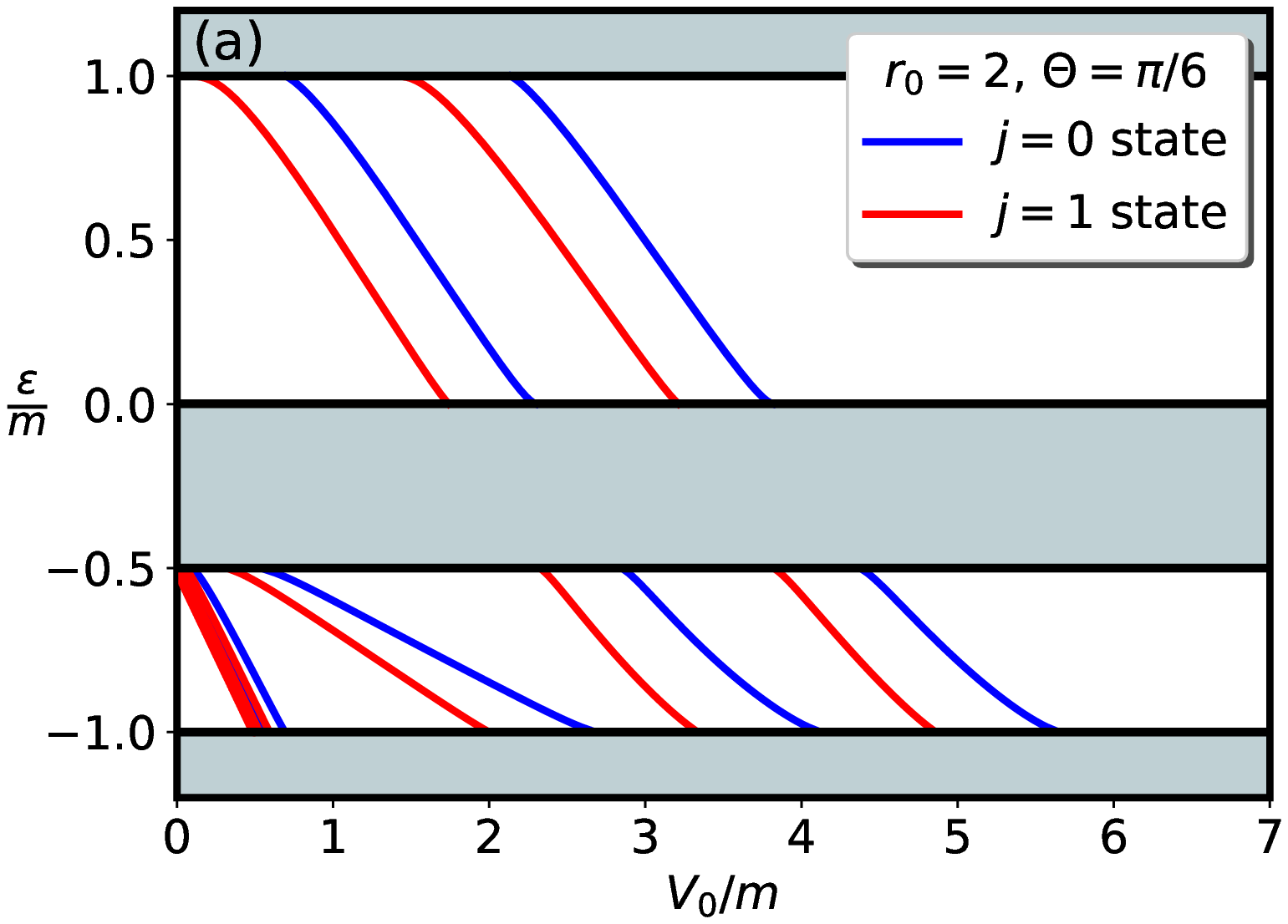}
	\includegraphics[scale=0.45]{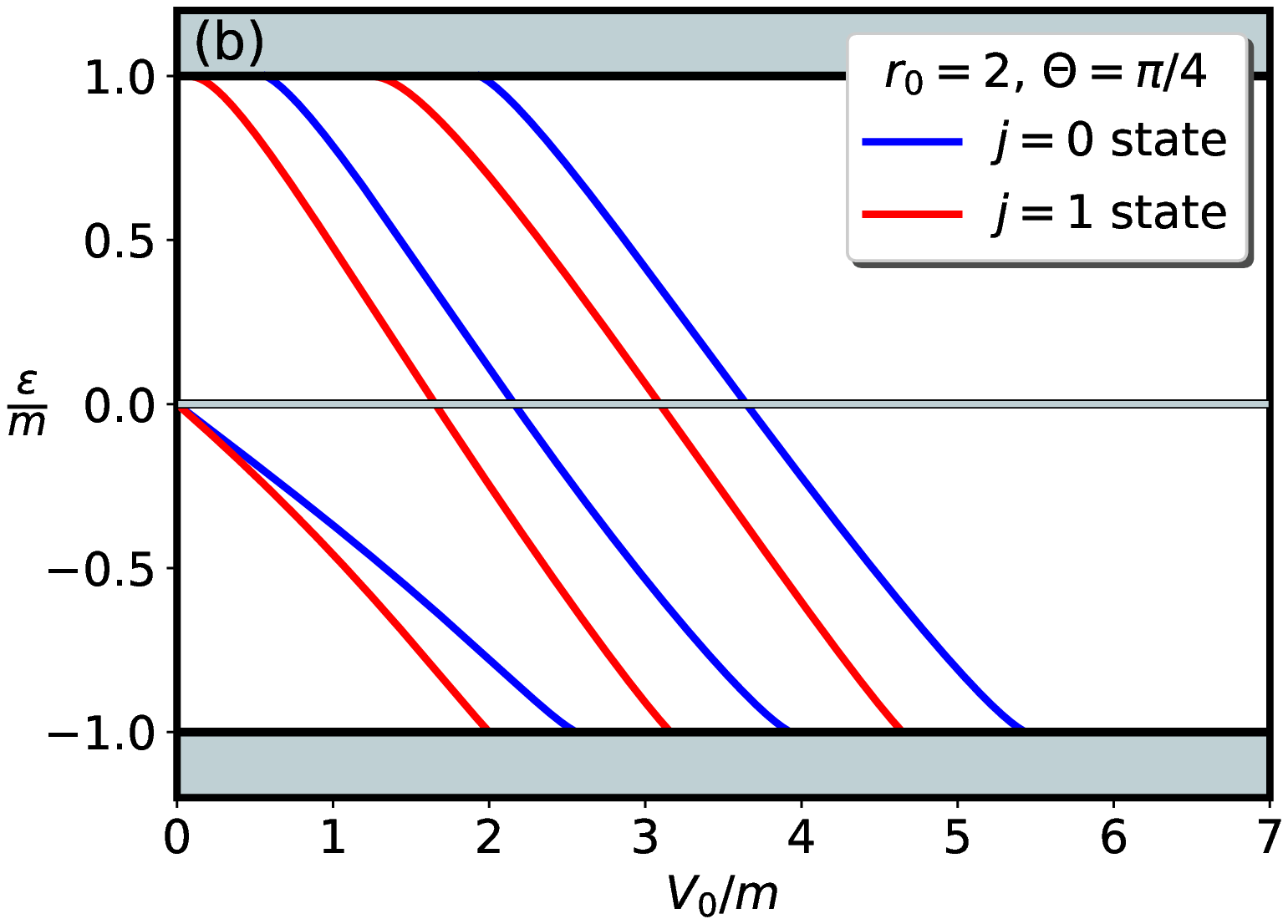}
	\includegraphics[scale=0.45]{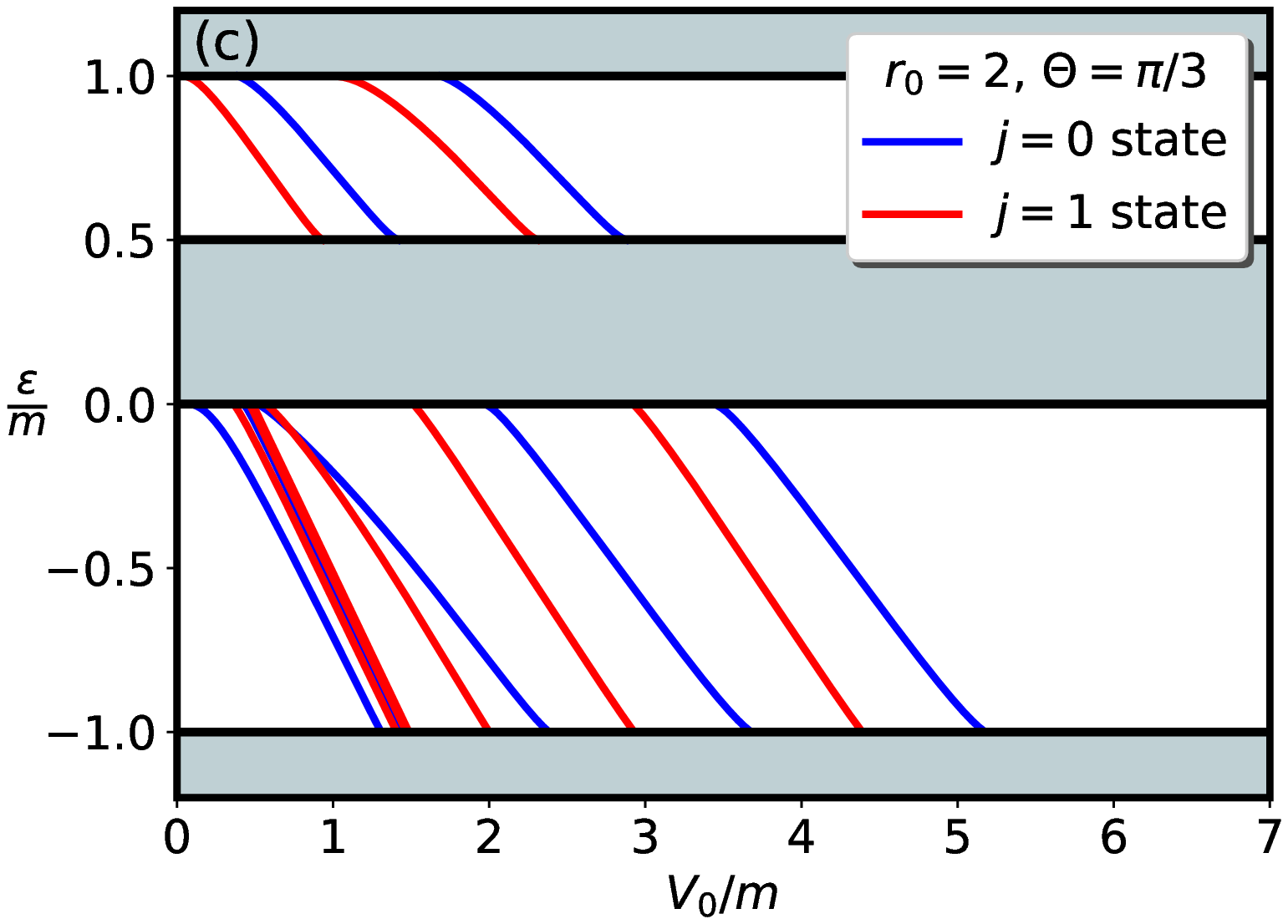}
	\includegraphics[scale=0.45]{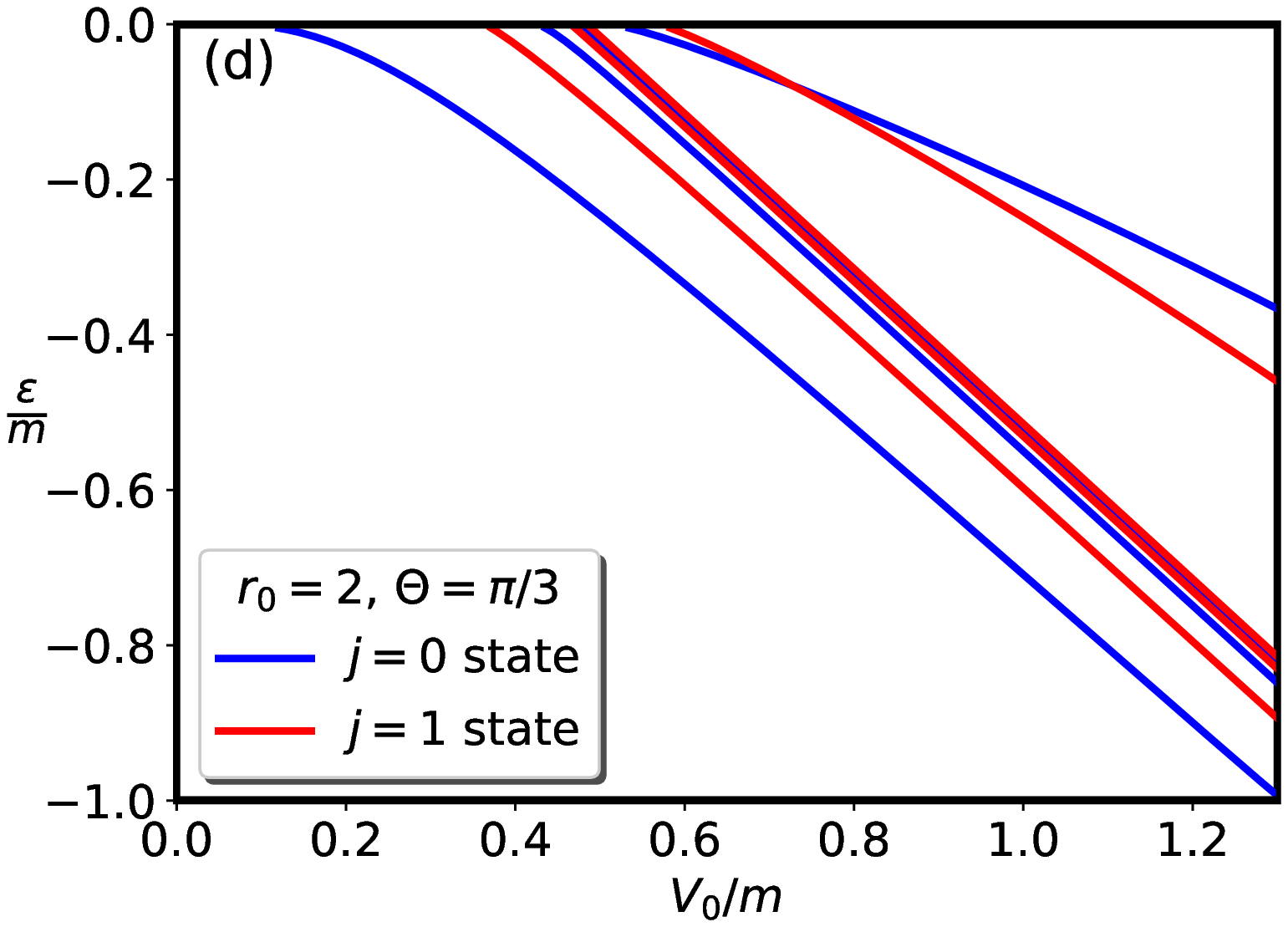}
	\caption{The spectrum of the first several bound states with the angular momentum $j=0$ and $j=1$ for the three different value
	$\Theta=\pi/6, \pi/4, \pi/3$. Panel (d) is a zoom-in of (c) panel for small values of the potential strength $V_0$ and the
	lower spectral gap.}		
	\label{fig:potential_well}
	\end{figure}

In order to get a general picture of the evolution of bound state solutions and compare the
corresponding results with those obtained analytically above, we numerically solve
Eq.\eqref{spectrum-1} and plot $\epsilon/m$ as a function of $V_0$ for the first several solutions with $j=0,1$ in
Fig.\ref{fig:potential_well} at the three representative values $\Theta=\pi/6, \pi/4, \pi/3$.
The behavior of the spectra near boundaries of the upper, middle, and lower bands agrees very well with our analytical results
presented above. According to panels (a) and (c), the general picture of the evolution of bound states with $V_0$ is the following.
Starting from some non-zero critical value of $V_0$, bound state solutions split from the upper continuum $\epsilon=m$, traverse the upper
spectral gap, dive into the middle band, and then reappear near the lower boundary of the middle band. Obviously, the states diving to and
splitting from the middle band can be smoothly connected by drawing the corresponding imaginary lines. This conclusion is especially evident
from panel (b) where the evolution of bound state solutions is shown in the dice model whose middle band is completely flat.
Finally, as $V_0$ increases further, these states dive into the lower continuum $\epsilon=-m$.

According to panels (a)-(c), there are also bound states which split from the middle band at any small $V_0$. As the potential
strength $V_0$ increases, these states dive into the lower continuum. A red strip observed on panels (a) and (c)
represents, in fact, a visual merging of distinct bound state solutions. A zoom-in of panel (c) for small values of $V_0$
plotted in panel (d) shows this explicitly. The two highest blue and red curves on this zoom-in demonstrate the level crossing
of solutions with $j=0$ and $j=1$. Since $j$ is different for these states, such a crossing is not forbidden by the von Neumann--Wigner theorem
[\onlinecite{Neumann-Wigner}].

\section{Coulomb center}
\label{section:center}

In this section, we will find bound state solutions for pseudospin-1 fermions in the Coulomb field of a charged impurity. We begin
with the system of equations \eqref{eq:main-system-1}-\eqref{eq:main-system-3} setting there $V(r)=-Z\alpha/r$.

We showed in Sec.\ref{section:model} that the system of the first order differential equations
(\ref{eq:main-system-1}) - (\ref{eq:main-system-3}) leads to relation (\ref{constraint-equation}) among the spinor components
$a,b,c$ which does not contain derivatives. Such a relation is important because it allows us to obtain a second order differential
equation for one component, e.g., $c(r)$ rather than a third order differential equation as one would naively expect.
Unfortunately, the corresponding equation, unlike Eq.(\ref{dice-model-equation-c}) in the case of a potential well, does not admit,
in general, solutions in terms of the known special functions. Still analytic solutions can be found in some particular cases.

\subsection{Some analytic solutions}

In this subsection we determine analytic bound state solutions for pseudospin-1 fermions in the field of a charged impurity by using
power series expansion and analyze analytically the presence of zero energy solutions in the dice model $(\Theta=\pi/4)$ for which
the system of \eqref{eq:main-system-1}-\eqref{eq:main-system-3} simplifies significantly.

\subsubsection{Expansion in series}

We can seek solutions for bound states of the system
(\ref{eq:main-system-1}) - (\ref{eq:main-system-3}) by using the power series expansion for spinor components of the form
\begin{align}
a(\rho)=e^{-\rho}\rho^{s}\sum\limits_{\nu=0}^{\infty}\rho^{\nu}a_{\nu},\quad	b(\rho)=e^{-\rho}\rho^{s}\sum\limits_{\nu=0}^{\infty}\rho^{\nu}b_{\nu},\quad 	 c(\rho)=e^{-\rho}\rho^{s}\sum\limits_{\nu=0}^{\infty}\rho^{\nu}c_{\nu}.
\label{solutions-series}
\end{align}
Clearly, the behavior of the wave function at infinity is determined by $\mbox{exp}[-\rho]$ and the expansion proceeds over
$\rho=vr$ with $v=\sqrt{\frac{\epsilon(m^2-\epsilon^2)}{\epsilon+m\cos2\Theta}}$. Then, we get the system of recursion relations for the
coefficients
$a_\nu,b_\nu,$ and $c_\nu$ which we write in the matrix form
\begin{equation}
M_1\Psi_{\nu-1}+M_2(\nu)\Psi_{\nu}=0,
\label{series-expansion}
\end{equation}
where $\Psi^T_{\nu}=(a_\nu,b_\nu,c_\nu)$ and the matrices $M_1,M_2$ are
\begin{eqnarray}
M_1=\left(\begin{array}{ccc}
\frac{m-\epsilon}{v} & -\cos\Theta & 0\\
-\cos\Theta & \frac{\epsilon}{v} & -\sin\Theta\\
0 & -\sin\Theta & -\frac{m+\epsilon}{v}
\end{array}\right),\quad
M_2(\nu)=\left(\begin{array}{ccc}
-Z\alpha & \cos\Theta (\nu+s+j-1) & 0\\
\cos\Theta (\nu+s-j) & Z\alpha & \sin\Theta (\nu+s+j)\\
0 & \sin\Theta (\nu+s-j-1) & -Z\alpha
\end{array}\right).
\label{matrices-M1-M2}
\end{eqnarray}
Setting $\nu=0$ in Eq.(\ref{series-expansion}) and taking into account that $\Psi_{-1}\equiv0$ we find that a nontrivial solution
for zero components $a_0,\,b_0,\,c_0$ exists if $\text{det}[M_2(0)]=0$ which determines the exponent $s$ and, thus, the behavior of
solutions at small distances
\begin{align}\label{eq:scale_factor}
	s=\frac{1}{2}+\sqrt{\frac{1}{4}+j^2-j\cos2\Theta-(Z\alpha)^2}
\end{align}
[compare with a similar solution in the dice model in [\onlinecite{Sun}]].
The second solution with the negative sign before the square root in the equation above leads to a divergent wave function at
$\rho=0$ and should be rejected. However, when the expression under the root becomes negative for a charge $Z$ exceeding some
critical value $Z_c$, both exponents are admissible and the ``fall-to-the-center" takes place for pseudospin-1 fermions. The
situation is the same as for the
three and two dimensional fermions described by the corresponding Dirac equations [\onlinecite{fall-to-center}]. As is well known,
the fall-to-the-center is avoided if the singular $1/r$ potential is replaced by a regularized one
(in the condensed matter setting the Coulomb potential is always naturally regularized at least by the lattice size). For subcritical
charges, $Z<Z_c$, the spectrum in the case of the non-regularized Coulomb potential could be determined if the series in Eq.
(\ref{solutions-series}) terminate at some $\nu=N$.

\begin{figure}
\centering
\includegraphics[scale=0.6]{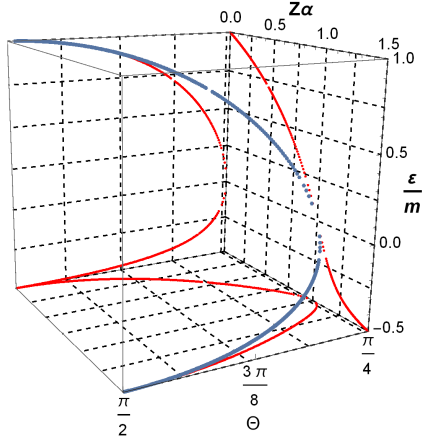}
\caption{The values of $\epsilon$, $Z\alpha$, and $\Theta$ which admit solutions of Eqs.(\ref{relations13}) for $j=1$ (blue curve). The
projections of the blue curve onto three planes are marked by red curves.
}
\label{quasiexact-solutions}
\end{figure}

To find possible terminations we note that the determinant of the matrix $M_1$ is zero in view of the definition of $v$.
Further, $\text{det}[M_2(\nu)]=Z\alpha\nu(2s+\nu-1)$ is nonzero for $\nu\ge1$, where we used the definition of $s$. Therefore, the
system of equations (\ref{series-expansion}) makes possible to determine $\Psi_\nu$ in terms of $\Psi_{\nu-1}$. The condition that
the series terminate at certain $N$ leads to the two following systems of equations:
\begin{align}
	M_{2}(0)\Psi_{0}=0,\quad \Psi_{N+1}=M(N)\Psi_{0}=0, \quad M(N)=(-1)^{N+1}\prod\limits_{l=N+1}^{1} \left[M_{2}^{-1}(l)M_{1}\right],\quad
	N\ge0.
\end{align}
Clearly, there are six equations for the three variables $a_0,c_0,b_0$. However, because $\text{det}[M_2(0)]=0$ and
$\text{det}[M(N)]=0$, there are, in fact,
four independent equations for three variables. For such a system of linear equations to be consistent, it is necessary that its rank to be less
than three. For example, for the termination $N=0$, the minors of the third rank must vanish, and we obtain the three conditions for the
existence of a nontrivial solution
\begin{align}
\label{relations13}
s+j-1=Z\alpha\frac{v}{m-\epsilon},\quad v\left((s-j)\frac{\cos^2\Theta}{m-\epsilon}-(s+j)\frac{\sin^2\Theta}{m+\epsilon}\right)+Z\alpha=0,\quad
s-j-1=-Z\alpha \frac{v}{m+\epsilon}.
\end{align}
Only two of them are independent. For example, the second equation is a linear combination of the first and third equations.
For the termination $N=1$, we have $\Psi_{2}=0$ and the following two systems of equations:
\begin{align}\label{N1equations}
	M_{2}(0)\Psi_0=0,\quad M_{1}M_{2}^{-1}(1)M_{1}\Psi_0=0.
\end{align}
Since solutions with $M_{2}(0)\Psi_0=0$ and $M_1\Psi_0=0$ belong to the termination $N=0$, we looks for solutions of the above
matrix equations with $M_1\Psi_0 \ne 0$. For terminations with $N \ge 1$, the system of equations (\ref{N1equations}) always gives two
independent conditions to determine eigenenergies. One of them can be derived straightforwardly from the system (\ref{series-expansion})
as a consequence of $\text{det}M_1=0$. Multiplying the first equation of the system  by $v \cos\Theta /(m-\epsilon)$,
the third equation by $-v\sin\Theta /(m+\epsilon)$ and adding them to the second equation, we find the relation between
$a_{\nu},\,b_{\nu}$ and $c_{\nu}$ valid at arbitrary $\nu$
\begin{align}
\label{eq:abc-relation}
&a_{\nu } \cos \Theta \left( -j+\nu +s-\frac{\alpha  v Z}{m-\epsilon}\right)+ b_{\nu } \sin \Theta \left( j+\nu +s+\frac{\alpha  v Z}{m+\epsilon}\right)+\nn
&c_{\nu } \left(\frac{v \sin ^2 \Theta  (j-\nu -s+1)}{m+\epsilon }+\frac{v \cos ^2 \Theta(j+\nu +s-1)}{m-\epsilon }+\alpha  Z\right)=0.
\end{align}
Next, we set $a_{N+1}=b_{N+1}=c_{N+1}=0$. Then we find from $M_1\Psi_N=0$ that
\begin{align}
	a_{N}=\frac{v\cos\Theta}{m-\epsilon}c_{N},\quad b_{N}=-\frac{v\sin\Theta}{m+\epsilon}c_{N}.
\end{align}
Together with Eq.(\ref{eq:abc-relation}) at $\nu=N$ this gives a nontrivial solution for $c_N$ when the following condition is
satisfied:
\begin{align}
v (2 N+2 s-1) (m \cos2\Theta+\epsilon )^2+Z\alpha\left(m \cos (2 \Theta\right )\left(m^2-3 \epsilon ^2\right)-2 \epsilon ^3)=0.
\label{spectral-equation-Coulomb-center}
\end{align}
One can check that for $N=0$ this equation coincides with the second equation in system (\ref{relations13}) if $s+j$ and $s-j$ are
substituted by their expressions from the first and third equations, respectively. We did not manage to find the second relation among
$a_N,c_N,b_N$ in a convenient form at arbitrary $N$. Unfortunately, their explicit expressions become very cumbersome with
increasing $N$. In any case, these two conditions give analytical solutions for a countably infinite set of values of charge
$Z\alpha$ at fixed angle $\Theta$. Thus, the considered system presents an example of the so-called quasiexactly solvable model.
Some selected solutions for $N=0$, $j=1$, and $\pi/4<\Theta<\pi/2$ fall on the blue curve shown in Fig.\ref{quasiexact-solutions}.

For $\Theta=0$ and $\pi/2$, we deal with graphene-like system and the problem greatly simplifies since the recursive relations
(\ref{series-expansion}) contain in this case only two equations for two coefficients. For example, for $\Theta=0$, the matrices $M_1$ and
$M_2(\nu)$ take the form
\begin{eqnarray}
M_1=\left(\begin{array}{cc}\frac{m-\epsilon}{u}&-1\\-1&\frac{\epsilon}{u}\end{array}\right),\quad M_2(\nu)=
\left(\begin{array}{cc}-Z\alpha&\nu+s+j-1\\ \nu+s-j&Z\alpha\end{array}\right),
\end{eqnarray}
where now $u=\sqrt{\epsilon(m-\epsilon)}$ and $s=1/2+\sqrt{(j-1/2)^2-(Z\alpha)^2}$. We find the spectrum
\begin{eqnarray}
\epsilon_{n,j}=\frac{m}{2}\left[1+\left(1+\frac{(Z\alpha)^2}{(n+s-1/2)^2}\right)^{-1/2}\right],\quad \left\{\begin{array}{c}n=0,1,2,\dots,\,j\ge1,\\
n=1,2,\dots,\,j\le0.\end{array}\right.
\end{eqnarray}
Under the change $l\rightarrow j+1/2, \epsilon\rightarrow (\epsilon+m)/2$ the spectrum turns into that for a Coulomb center in
graphene [\onlinecite{graphene-Coulomb}].

\subsubsection{The absence of solutions with zero energy in the dice model}

Perhaps one of the most interesting cases is the presence of zero energy solutions and the fate of the flat band in the dice model in the
field of the Coulomb potential. Our analysis in Subsec.\ref{sec:well-solutions} and Appendix
\ref{sec:dice} shows that the flat band survives in a potential well although there are electron bound states which split from the flat band
even in the case of an arbitrary shallow potential well. In order to see whether the flat band survives in the Coulomb center, we consider the
system of equations \eqref{eq:main-system-1}-\eqref{eq:main-system-3} which for $\Theta=\frac{\pi}{4}$ and $\epsilon=0$ simplifies to
\begin{align}
\label{flat-band-system}
&\frac{1}{\sqrt{2}}\left(c'+\frac{j-1}{\rho}c\right)+\left(1-\frac{Z\alpha}{\rho}\right)a=0,\nonumber\\
&\frac{1}{\sqrt{2}}\left(a'-\frac{j}{\rho}a\right)+\frac{1}{\sqrt{2}}\left(b'+\frac{j}{\rho}b\right)+\frac{Z\alpha}{\rho}c=0,\\
&\frac{1}{\sqrt{2}}\left(c'-\frac{j+1}{\rho}c\right)-\left(1+\frac{Z\alpha}{\rho}\right)b=0,\nonumber
\end{align}
where $\rho=mr$. By adding and subtracting the first and third equations, we obtain
\begin{align}
\label{flat-band-system-1}
 	\sqrt{2}\left(c'-\frac{1}{\rho}c\right)+(a-b)-\frac{Z\alpha}{\rho}(a+b)=0,\quad
  Z\alpha(a-b)=\sqrt{2}j c+(a+b)\rho.
\end{align}
The latter equation makes possible to eliminate $a-b$ from the second equation in (\ref{flat-band-system}) and the first equation in
(\ref{flat-band-system-1}). Then we find
\begin{align}
	\label{eq:sum_without_a-b}
	&\frac{1}{\sqrt{2}}(a'+b')-\frac{j}{\sqrt{2}\rho}\frac{\sqrt{2}j c+(a+b)\rho}{Z\alpha}+\frac{Z\alpha}{\rho}c=0,\nonumber\\
	&\sqrt{2}\left(c'-\frac{1}{\rho}c\right)+\frac{\sqrt{2}j c+(a+b)\rho}{Z\alpha}-\frac{Z\alpha}{\rho}(a+b)=0.
\end{align}
Determining $c$ from the first equation of the system above and substituting it in the second equation, we obtain the following second order
differential equation for $Q=a+b$:
\begin{align}
	Q''-\bigg[1+\frac{j^2-Z^2\alpha^2}{\rho^2}\bigg]Q=0,
\end{align}
whose general solution can be easily expressed in terms of the modified Bessel functions
\begin{align}
	Q=\sqrt{\rho}\left[C_1 I_{\nu}(\rho)+C_2 K_{\nu}(\rho)\right], \quad \quad
	\nu=\sqrt{1/4+j^2-Z^2\alpha^2}.
\end{align}
Since $I_{\nu}(\rho)$ is an exponentially growing function at large $\rho$, we should set $C_1$ to zero in order to have normalized solutions.
Further, since $K_{\nu}(z) \sim z^{-\nu}$ at small $z$, we should set $C_2=0$ unless $Z\alpha$ is exactly equal to $|j|$ (the corresponding
solutions will be analyzed in the next subsection). Since there are no solutions with zero energy, we conclude that the flat band in the dice
model {\sl does not survive} in the pure Coulomb potential like it happens in the case of Landau levels for the electrons
in the Coulomb field. The zero-energy solutions do not exist also for a regularized Coulomb potential of the form
$V(\rho)=-Z\alpha\left(\theta(\rho_0-\rho)/\rho_0+\theta(\rho-\rho_0)/\rho\right)$. Indeed, though in this case we can match two solutions,
$I_{\nu}(\rho)$ and $ K_{\nu}(\rho)$ at the boundary $r=r_0$ (Eq.\ref{EOM-2-condition}), we cannot satisfy the boundary condition
(\ref{EOM-1-condition}). The numerical studies in the next section confirm the absence of zero-energy solution for a regularized Coulomb
potential. This result is different from that in the case of a potential well where, according to the analysis in Appendix \ref{sec:dice}, the
flat band survives in a potential well. Technically, the reason is that Eq.(\ref{dice-equation-c-1}) is automatically satisfied outside the
potential well. Therefore, the solution is arbitrary there and could be always matched to that in the potential well.

\subsection{Numerical solutions}

\begin{figure}
	\centering
	\includegraphics[scale=0.33]{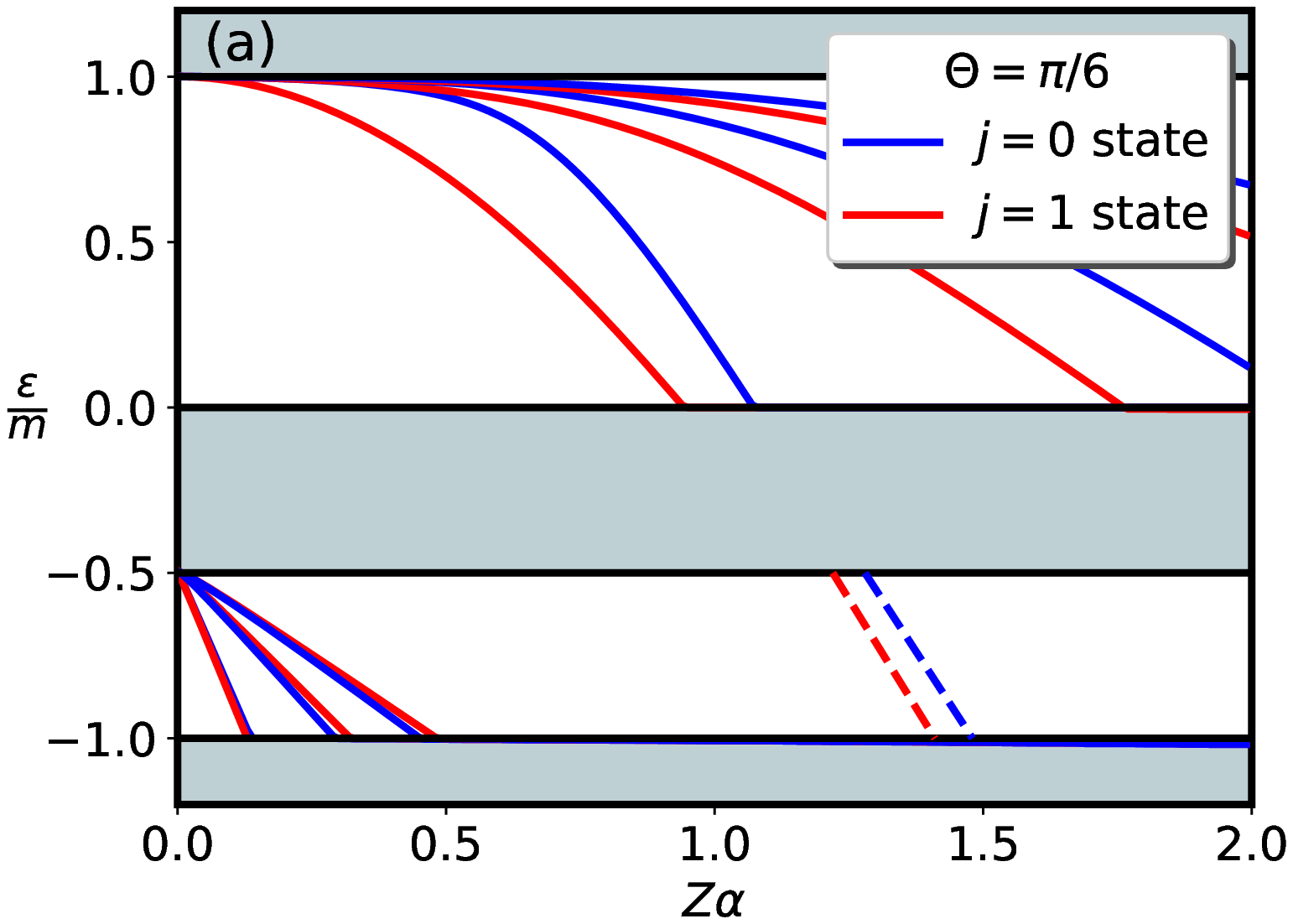}
	\includegraphics[scale=0.33]{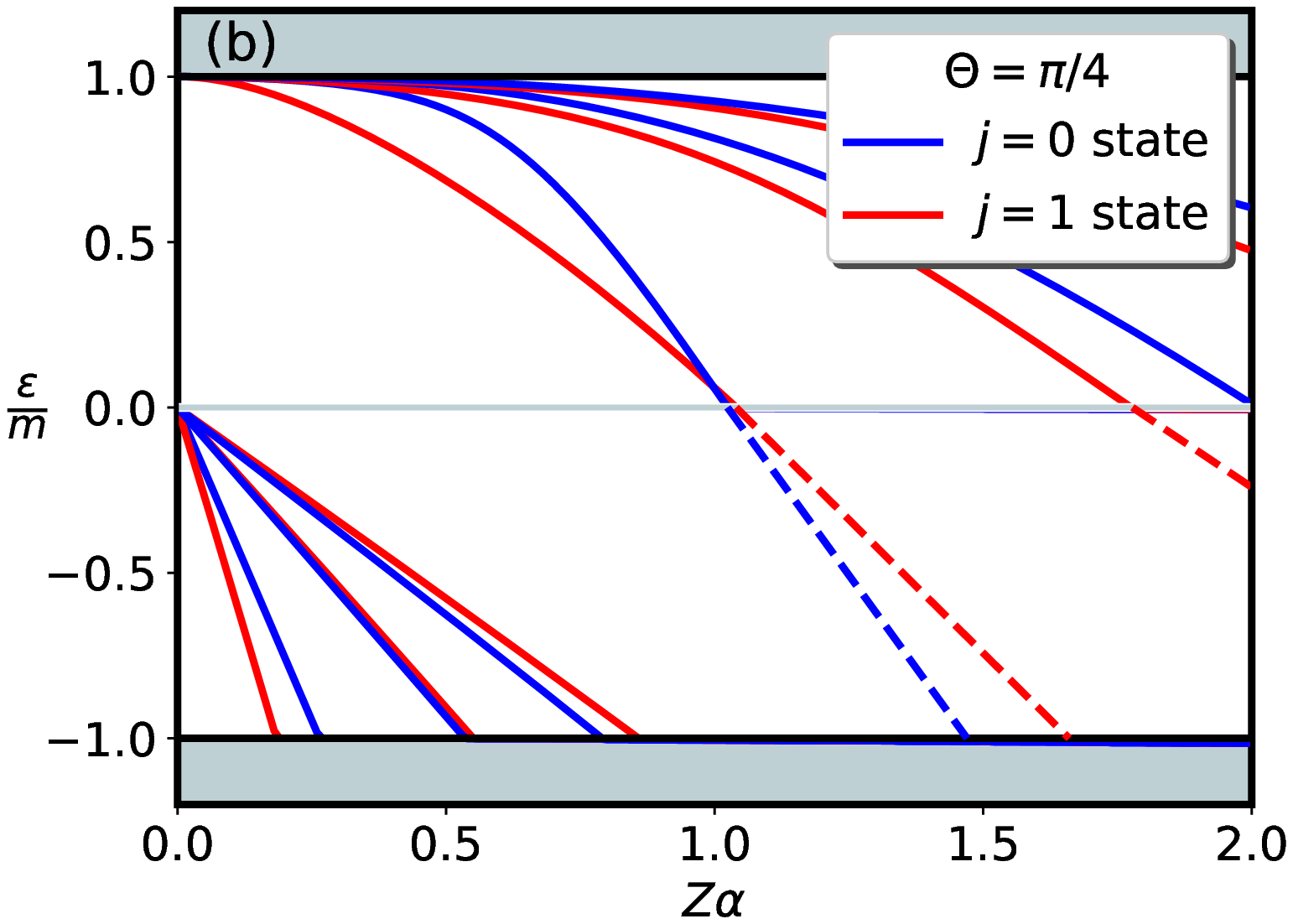}
	\includegraphics[scale=0.33]{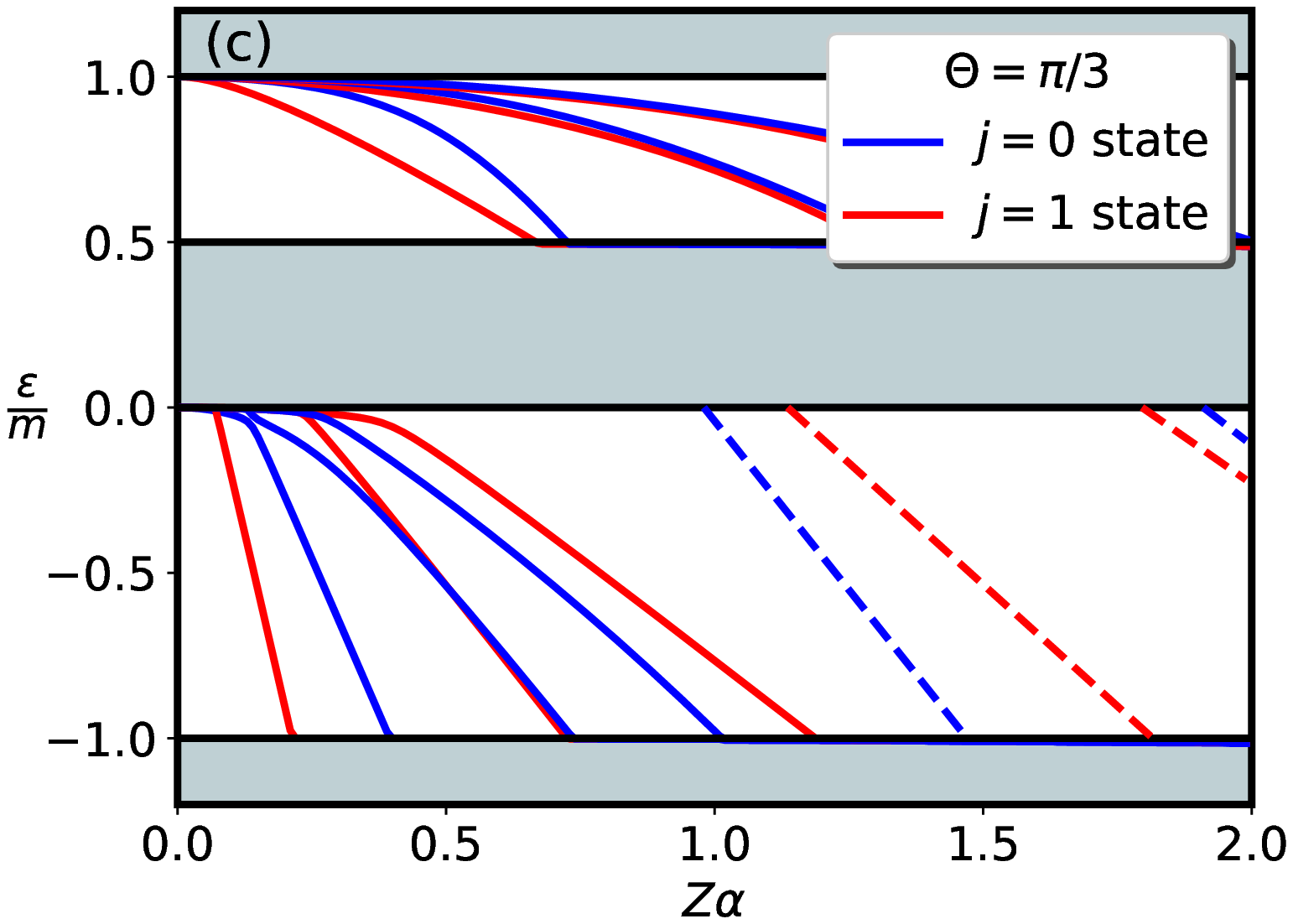}
	\caption{Numerically solutions of the system of equations \eqref{system:integral_equations} for
	$\Theta=\frac{\pi}{6}, \frac{\pi}{4}, \frac{\pi}{3}$ and $\Lambda/m=10$.}
	\label{fig_numerical}
\end{figure}

Since there are no analytical solutions for the spectrum in the case of arbitrary $\Theta$ and $Z\alpha$,
below we determine bound state solutions for pseudospin-1 fermions in a regularized Coulomb potential numerically.
For this, we adopt the same approach as in Ref.[\onlinecite{Oriekhov}] and solve the following integral equation in momentum space:
\begin{align}
H_0(\mathbf{k})\Psi(\mathbf{k})+\int^\Lambda\frac{d^2 q}{(2\pi)^2}\Psi(\mathbf{q})V_{reg}(\vec{k}-\vec{q})=\epsilon\Psi(\mathbf{k}),
\end{align}
where $H_0(\mathbf{k})$ is given by Eq.(\ref{Hamiltonian-free}) and
\begin{align}
&V_{reg}(\mathbf{k})=-\frac{2\pi \zeta}{|\mathbf{k}|},\quad \Psi(\mathbf{k})=\left(\begin{array}{c}
a_{j}(k)e^{i(j-1)\phi}\\
c_{j}(k)e^{ij\phi}\\
b_{j}(k)e^{i(j+1)\phi}
\end{array}\right).
\end{align}
Here $\Lambda$ is the wave vector cut-off and $\zeta=Z\alpha / \kappa$ in the regularized Coulomb potential includes the dielectric constant
$\kappa$ of substrate. The system for the components $a,c,b$ of the wave function has the form
\begin{align}
& k \cos\Theta c_{j}(k)+m a_{j}(k)-\frac{\zeta}{2\pi}\int_{0}^{\Lambda}\,dq q a_j(q) \int_{0}^{2\pi}
\frac{d\alpha}{|\mathbf{k}-\mathbf{q}|}e^{i(j-1)(\alpha-\phi)}=\epsilon a_{j}(k),\\
&k \cos\Theta a_{j}(k)+k \sin\Theta b_{j}(k)-\frac{\zeta}{2\pi}\int_{0}^{\Lambda}\,dq q c_j(q) \int_{0}^{2\pi}
\frac{d\alpha}{|\mathbf{k}-\mathbf{q}|}e^{ij(\alpha-\phi)}=\epsilon c_{j}(k),\\
&k \sin\Theta c_{j}(k)-m b_{j}(k)-\frac{\zeta}{2\pi}\int_{0}^{\Lambda}\,dq q b_j(q) \int_{0}^{2\pi}
\frac{d\alpha}{|\mathbf{k}-\mathbf{q}|}e^{i(j+1)(\alpha-\phi)}=\epsilon b_{j}(k).
\end{align}
The integrals over the angle $\alpha$ can be calculated analytically [\onlinecite{Oriekhov}],
\begin{align}
K_{j}(k,q)=\int_{0}^{2\pi}\frac{\cos(j\alpha)d\alpha}{\sqrt{k^2+q^2-2kq\cos\alpha}}=\frac{2}{\sqrt{kq}}Q_{|j|-1/2}
\left(\frac{k^2+q^2}{2kq}\right),
\end{align}
where $Q_{\nu}(z)$ is the Legendre function of the second kind. Rewriting the above system of equations in terms of kernels $K_{j}$, we obtain
\begin{align}
& k\cos\Theta  c_{j}(k)+m a_{j}(k)-\frac{\zeta}{2\pi}\int_{0}^{\Lambda}\,dq q K_{j-1}(k,q) a_j(q) =\epsilon a_{j}(k),\nn
&k \cos\Theta a_{j}(k)+ k \sin\Theta b_{j}(k)-\frac{\zeta}{2\pi}\int_{0}^{\Lambda}\,dq q K_{j}(k,q) c_j(q) =\epsilon c_{j}(k),\\
\label{system:integral_equations}
&k \sin\Theta c_{j}(k)-m b_{j}(k)-\frac{\zeta}{2\pi}\int_{0}^{\Lambda}\,dq q K_{j+1}(k,q) b_j(q) =\epsilon b_{j}(k).\nonumber
\end{align}
Further, we replace the integrals in the above system of equations by regularized sums
\begin{align}
\int_{0}^{\Lambda} dq q K_{j}(k,q) c_{j}(q)=\sum\limits_{i=0}^{N} q_i w_i K_{j}(k, q_i)[c_j (q_i)-c_j(k)]+c_j(k)\int_{0}^{\Lambda} dq q
K_{j}(k,q).
\end{align}
The analytical expressions for the last integral can be found in Ref.[\onlinecite{Oriekhov}]. We solved
numerically the system of equations (\ref{system:integral_equations}) for the states with different $j$, the corresponding results for the
states $j=0,1$ are presented in Fig.\ref{fig_numerical}.
As expected, the energy levels monotonously decrease as $Z\alpha$ increases and then dive into the middle or lower continua. For the clarity of
presentation, only few levels are plotted in the lower spectral gap. In particular, the dashed lines correspond to
bound states which are the continuation of bound state levels from the upper spectral gap (this is especially clear from panel (b) in the case
of the dice model). This picture is in agreement with results for the potential well in Sec.\ref{sec:well-solutions}. We found also
that if the cut-off is sufficiently large, then the electron bound state with $j=0$ crosses the state $j=1$ in the upper spectral gap as the
charge of impurity increases. For intermediate values of a cut-off like in Fig.\ref{fig_numerical}, such a crossing maybe absent.  On the other
hand, no such a crossing is observed in the lower spectral gap between the central and lower continua.

\section{Summary}
\label{section:summary}

We studied the electron states of pseudospin-1 fermions of the $\alpha-{\cal T}_3$ lattice with a gap term in the field of a charged
impurity. We analyzed electron bound states in the two cases:  a long-range Coulomb potential (both regularized and unregularized) and the
short-range centrally symmetric potential well.  In the later case, matching solutions inside and outside the potential well, the spectral
equation for the electron bound states is found. The analytical results are presented for the energy levels near continuum
boundaries. The analysis shows that a critical value of the strength of the potential $V_{0,cr}$ is needed
in order that bound states split from the upper continuum. For sufficiently large $V_0$, such states dive into the central continuum. One of
our principal findings is that electron bound states split from the central continuum at arbitrary small $V_0$.

The same result applies to the dice model, which is a particular case of the $\alpha-{\cal T}_3$ model with $\Theta=\pi/4$. Its central band
is a degenerate completely flat band. This band survives in a centrally symmetric potential well, although as the strength of the
potential increases, more and more bound states with higher value of the total angular momentum split from the flat band.  The main reason
responsible for the survival of the flat band in a sharp potential well is the fact that the solution
outside the potential well is an arbitrary function and, thus, can always be matched to the solution inside the potential well. Consistent with
this, we found that in the Coulomb potential, where the region without the potential perturbation is absent, there are no solutions with zero
energy. Consequently, the flat band does not survives in this case. The Coulomb potential is the long-range one, while the sharp potential well
corresponds to the case of a  potential with compact support. These two potentials are, in a certain sense, two opposite
cases. Therefore, it would be interesting to study whether the flat band survives in the case of a short-range potential with noncompact support.

For the genuine Coulomb potential, by using a generalized expansion in a power series, we
were able to obtain analytic solutions for a countable set of values of impurity charge $Z\alpha$ at fixed $\Theta$. For a regularized
Coulomb potential, we found electron bound states by solving numerically the integral equation in momentum space.
The bound states solutions descend from the upper continuum $\epsilon=m$ and the
central continuum ($\epsilon=-m\cos\Theta$ for $0<\Theta<\pi/4$ and $\epsilon=0$ for $\pi/4<\Theta<\pi/2$). As the charge of impurity increases,
these states dive into the central and lower continuum, respectively, at certain critical charges. We found that the state with the total
angular momentum $j=1$ first splits from the upper continuum. The electron bound state with $j=0$ may cross the state $j=1$
in the upper spectral gap as the charge of impurity increases. On the other hand, no such a crossing is
observed in the lower spectral gap between the central and lower continua.

It is well known that the atomic collapse in the one electron problem in the field of a charged impurity is a precursor of the
supercritical instability and the quasiparticle gap generation in a many body system [\onlinecite{Gamayun2010}]. Therefore, a natural extension
of the present study would be the investigation of the gap generation for pseudospin-1 fermions in the $\alpha-{\cal T}_3$ model.

\section{Acknowledgments}

We are grateful to Jean-Noel Fuchs for fruitful discussions on the topological properties of the considered model.
The work of E.V.G. and V.P.G. is partially supported by the National Academy of Sciences of Ukraine (project 0116U003191) and by its Program of
Fundamental Research of the Department of Physics and Astronomy (project No. 0117U000240). V.P.G. acknowledges the support of the RISE Project
CoExAN GA644076.

\appendix
\section{Solutions for $\Theta>\frac{\pi}{4}$ in a potential well}
\label{sec:large-angle}

Since solutions near the upper continuum $\epsilon \to m$ and the lower continuum $\epsilon\to -m$ are the same as found in
Sec.\ref{sec:well-solutions} for $\Theta<\frac{\pi}{4}$, we do not repeat the corresponding analysis here.

\subsection{Solutions diving into the middle gap}

We look for solutions $\epsilon=-m\cos 2\Theta+x$ with $x \to 0$. While the function $v_1(\epsilon+V_0)$ remains finite for $x\to0$,
the function $v_2(\epsilon)$ behaves as
\begin{align}
 v_2(\epsilon)\simeq m\sin2\Theta\sqrt{\frac{m|\cos 2\Theta|}{x}},\quad x\to0.
\end{align}

For {\bf the state $j=0$}, the spectral equation \eqref{spectrum-2a} takes the form
\begin{align}
	\frac{V_0}{(m|\cos 2\Theta|+V_0)^2-m^2}\frac{v_1(\epsilon+V_0)J_1(v_1(\epsilon+V_0)r_0)}{J_0(v_1(\epsilon+V_0)r_0)}
	=-\sqrt{\frac{x|\cos 2\Theta|}{m\sin^2 2\Theta}}.
\label{eq:j=0middleband}
\end{align}
The critical potential is determined by the first root of the Bessel function $J_1$, namely,
$v_1(m|\cos 2\Theta|+V_0)r_0=j_{1,1}$.
By using the identity $J_{1}^{\prime}(j_{1,1})=J_0(j_{1,1})$, and near the first root of $J_1$ Eq.(\ref{eq:j=0middleband}) and we find
the behavior of the energy level near the critical potential strength $V_0\lesssim V_{0,cr}$,
\begin{align}
	x=\frac{m\sin^2 2\Theta}{|\cos 2\Theta|}a^2 \bigg[(V_{0,cr}-V_0)r_0\bigg]^2,\quad a=\frac{V_{0,cr}}{(m|\cos 2\Theta|+V_{0,cr})^2-m^2}
v_1(m|\cos 2\Theta|+V_{0,cr})v_1'(m|\cos 2\Theta|+V_{0,cr}).
\end{align}

The spectral equation \eqref{spectrum-1} for {\bf the state $j=1$} becomes
\begin{align}
	\left(\frac{\cos^2 \Theta}{m(1-|\cos 2\Theta|)-V_0}J_0(v_1(V_0+\epsilon)r_0)+\frac{\sin^2 \Theta}{m(1+|\cos 2\Theta|)+V_0}
	J_2(v_1(V_0+\epsilon)r_0) \right)\frac{v_1(\epsilon+V_0)}{J_1(v_1(V_0+\epsilon)r_0)}\approx-\sqrt{\frac{x|\cos 2\Theta|}{m\sin^2 2\Theta}}.
\end{align}
Since $x \to 0$, the solution is defined by the root of the left-hand side of the above equation.

\subsection{Solutions descending from the middle gap}

For solutions descending from the central continuum from $\epsilon=0$, it is convenient to set $\epsilon=-x$ with $x>0$.

For {\bf the state $j=0$}, the functions $v_1(\epsilon+V_0)$ and $v_2(\epsilon)$ become
\begin{align}
	v_1(\epsilon+V_0)=\sqrt{\frac{(V_0-x)(m^2-(V_0-x)^2)}{m|\cos 2\Theta|-V_0+x}},\quad v_2(\epsilon)\simeq \sqrt{\frac{mx}{|\cos 2\Theta|}}.
\end{align}
The spectral equation \eqref{spectrum-2a} takes the form
\begin{align}
	\frac{V_0-m|\cos 2\Theta|}{V_0^2-m^2}\frac{v_1(V_0)J_{1}(v_1(V_0)r_0)}{J_{0}(v_1(V_0)r_0)}
	=\frac{|\cos 2\Theta|}{mr_0}\ln^{-1}\left(\frac{2e^{-\gamma}}{r_0}\sqrt{\frac{|\cos 2\Theta|}{mx}}\right).
\end{align}
The critical potential is lower than $m |\cos 2\Theta| $ and is determined as the first root of the Bessel function $J_{1}(v_1(V_0)r_0)$.

The equation \eqref{spectrum-1} for {\bf the state $j=1$} equals
\begin{align}
&\left(\frac{\cos^2\Theta }{m-V_0}J_{0}(v_1(V_0)r_0)
+\frac{\sin^2\Theta }{m+V_0}J_{2}(v_1(V_0)r_0)
\right)\frac{v_1(V_0)}{J_{1}(v_1 (V_0)r_0)}=\frac{2\sin^2\Theta}{m r_0}
\end{align}
and determines the critical value of potential for which the electron bound state $j=1$ separates from the central continuum. Since
$V_{0,cr}>0$, the state with $j=0$ is the lowest bound state.

\section{Solutions in dice model for $\Theta=\pi/4$}
\label{sec:dice}

The dice model is realized at $\Theta=\pi/4$. The central continuum in the $\alpha-{\cal T}_3$ model reduces in the dice model to a completely
flat degenerate band of zero energy $\epsilon=0$. As discussed in the Introduction, we recently found [\onlinecite{dice-model-boundaries}] that
the energy dispersion of the completely flat energy band of the dice model is not affected by the presence of boundaries except the trivial
reduction of the degenerated electron states due to the finite spatial size of the system. This gives rise to the question whether the electron
states of the flat band remain degenerate also in the presence of a potential perturbation. We study this question in this Appendix.

In the dice model with the potential well $-V_0\theta(r_0-r)$, Eq.(\ref{dice-model-equation-c}) leads to the following
equation for the $c$ component:
\be
\left(\epsilon+V_0\theta(r_0-r)\right)\left(c''-\frac{1}{r}c'-\frac{j^2-1}{r^2}c\right)-\left(\epsilon+V_0\theta(r_0-r)\right)
\left(m^2-(\epsilon+V_0\theta(r-r_0))^2\right)c=0.
\label{dice-equation-c-1}
\ee
The $a$ and $b$ components are defined through $c$ as follows:
\begin{equation}
a=\frac{c'+\frac{j-1}{r}c}{\sqrt{2}\,(\epsilon+V_0\theta(r-r_0)-m)},\quad\quad
b=\frac{c'-\frac{j+1}{r}c}{\sqrt{2}\,(\epsilon+V_0\theta(r-r_0)+m)}.
\label{dice-model-a-c}
\end{equation}
For $r<r_0$, we find the solution regular at $r=0$,
\begin{equation}
c(r)=c_1 r I_{|j|}(u_2(V_0)r),
\label{flat-band-solution-c}
\end{equation}
where $u_2(\epsilon)=\sqrt{m^2-\epsilon^2}$.

Now we should find solution at $r>r_0$. First of all,
we see that Eq.(\ref{dice-equation-c-1}) is identically satisfied for $r>r_0$ at $\epsilon=0$. This means that $c$ is an arbitrary function
in this region. [This can be seen directly also from system (\ref{flat-band-system}) setting $Z=0$, expressing
from the first and third equations of the system $a$ and $b$ through $c$, substituting them into the second equation, and checking that it
is automatically satisfied.]

In the absence of a potential, Eq.(\ref{spinor-spin1}) defines a solution at fixed momentum $\mathbf{k}$. Then the general solution
in coordinate space for the middle band in the dice model is defined as the Fourier transform of momentum space solutions with arbitrary
coefficient function $C(\mathbf{k})$
 \bea
 \psi_0(\mathbf{r})=\int \frac{d^2k}{(2\pi)^2}\frac{e^{i\mathbf{k}\mathbf{r}}\,C(\mathbf{k})}{\sqrt{2(\mathbf{k}^2+m^2})}\hspace{-1mm}
 \left(\hspace{-1mm}\begin{array}{c}k_x-i k_y\\
 -\sqrt{2}m\\ -(k_x+i k_y)\end{array}\hspace{-1mm}\right).
 \label{spinor-spin1-general}
 \eea
Further, in order to match this solution with the solution inside the potential well (\ref{dice-model-a-c}) and
(\ref{flat-band-solution-c}), we should determine
when the general solution (\ref{spinor-spin1-general}) defines the state with the angular momentum $j$.
According to Eq.(\ref{spinor-general-form-polar-coordinates}), the upper, middle, and lower components of such a state should depend on the
angular variable as $\mbox{exp}[i(j-1)\phi]$, $\mbox{exp}[ij\phi]$, and $\mbox{exp}[i(j+1)\phi]$, respectively. Let us check that the
natural choice $C(\mathbf{k})=A_j(k)\mbox{exp}[ij\phi_{\mathbf{k}}]$, where $A_j(k)$ is an arbitrary function of $k=|\mathbf{k}|$, provides the
necessary solution. Indeed, for the middle component, we have
\be
-\int^{+\infty}_{0}\int^{2\pi}_0 \frac{kdk\,d\phi_{\mathbf{k}}}{(2\pi)^2}\,\frac{m}{\sqrt{k^2+m^2}}\, A_j(k)\,e^{ij\phi_{\mathbf{k}}}
e^{i\mathbf{k}\mathbf{r}}
=-\int^{+\infty}_{0}\int^{2\pi}_0 \frac{kdk\,d\phi_{\mathbf{k}}}{(2\pi)^2}\,\frac{m}{\sqrt{k^2+m^2}}\, A_j(k)\,e^{ij\phi_{\mathbf{k}}}
e^{ikr\cos(\phi_{\mathbf{k}}-\phi)}.
\ee
By making the change of variable $\phi_{\mathbf{k}} \to \phi_{\mathbf{k}} + \phi$, we obtain
\begin{align}
-e^{ij\phi} \int\limits^{+\infty}_{0}\int\limits_{-\phi}^{2\pi-\phi} \frac{kdk\,d\phi_{\mathbf{k}}}{(2\pi)^2}\,\frac{m}{\sqrt{k^2+m^2}}\, A_j(k)\,e^{ij\phi_{\mathbf{k}}}
e^{ikr\cos(\phi_{\mathbf{k}})}=
-i^{j} e^{ij\phi} \int\limits^{+\infty}_{0} \frac{kdk}{2\pi}\,\frac{m}{\sqrt{k^2+m^2}}\, A_j(k) J_{j}(kr).
\end{align}
Here, in order to calculate the integral over $\phi_{\mathbf{k}}$, we took into account that this is the integral of a periodic
function over period. Therefore, the presence of $-\phi$ in the limits of integration is irrelevant. It is easy to check that the upper and
lower components of the spinor also result in the correct dependence on $\phi$.

Thus, we have the following general solutions with the total angular momentum $j$ in the
regions $r<r_0$ and $r>r_0$:
\begin{align}
	\Psi_{<} = c_1\left(\begin{array}{c}
	-\frac{\sqrt{m^2-V_0^2}}{\sqrt{2}(m-V_0)}J_{j-1}(\sqrt{m^2-V_0^2}r)e^{i(j-1)\phi}\\
	i J_{j}(\sqrt{m^2-V_0^2} r)e^{ij\phi}\\
	-\frac{\sqrt{m^2-V_0^2}}{\sqrt{2}(m+V_0)}J_{j+1}(\sqrt{m^2-V_0^2}r)e^{i(j+1)\phi}
	\end{array}\right), \quad \Psi_{>} = i^{j}\int\limits_{0}^{+\infty} \frac{kdk}{2\pi}\,
	\frac{A_j(k)}{\sqrt{2(k^2+m^2)}}\,\hspace{-1mm}
 \left(\begin{array}{c}
 ik J_{j-1}(kr)\,e^{i(j-1)\phi}\\
 \sqrt{2}m J_{j}(kr)\,e^{ij\phi}\\ -ikJ_{j+1}(kr)\,e^{i(j+1)\phi}\end{array}\hspace{-1mm}\right).
	\end{align}
According to Eqs.(\ref{EOM-1-condition}) and (\ref{EOM-2-condition}), the $c$ component and the sum $a+b$ of the $a$ and $b$ components of
the solutions inside and outside the potential well should be matched at $r=r_0$. Thus, we have
	\begin{align}
	&c_1 J_{j}(\sqrt{m^2-V_0^2}\,r_0)=i^{j-1}m\int\limits^{+\infty}_{0} \frac{kdk}{2\pi}\,
	\frac{m\,A_j(k) J_{j}(kr_0)}{\sqrt{k^2+m^2}},
	\label{matching-1}\\
	&c_1\left(	\frac{\sqrt{m^2-V_0^2}}{m-V_0}J_{j-1}(\sqrt{m^2-V_0^2}r_0)+\frac{\sqrt{m^2-V_0^2}}{m+V_0}J_{j+1}(\sqrt{m^2-V_0^2}r_0)\right)=\nn
	&-i^{j+1}\int\limits_{0}^{+\infty}\frac{k^2 dk}{2\pi}\frac{A_j(k)}{\sqrt{2(k^2+m^2)}}\bigg(J_{j-1}(kr_0)-J_{j+1}(kr_0)\bigg).
	\label{matching-2}
	\end{align}
Determining $c_1$ from Eq.(\ref{matching-1}) and substituting it into Eq.(\ref{matching-2}) results in one integral condition for
$A_j(k)$. This integral condition can be easily satisfied and practically does not restrict $A_j(k)$. This means the flat band does
{\it survive} in the presence of a potential well.

As to the bound states, the analytical formulas for the energy levels near the boundaries of upper and lower spectral gaps can be easily
obtained similarly to the analysis in the case $\Theta\neq0$. The corresponding results are presented in Fig.\ref{fig:potential_well}. They
show that an arbitrary weak perturbation $V_0$ leads to splitting of bound state solutions with $j=0$ and
$j=1$ from the flat band. Therefore, potential perturbations due to inevitable disorder will remove in real systems the absolute degeneracy
of the flat band in the dice model. We would like to note that disordered flat bands on the kagome lattice was recently studied in
[\onlinecite{Bilitewski}].

\end{document}